\documentclass[a4paper,fleqn,usenatbib]{mnras}

\usepackage{newtxtext,newtxmath}

\usepackage[T1]{fontenc}
\usepackage{ae,aecompl}

\usepackage{graphicx}	
\usepackage{amsmath}	
\usepackage{times}
\usepackage{graphics, subfigure}
\usepackage{epsfig}
\usepackage{multirow}
\usepackage{ulem}
\usepackage{pdfpages}

\def\simlt{\mathrel{\hbox{\rlap{\hbox{\lower4pt\hbox{$\sim$}}}\hbox{$<$}}}}
\def\simgt{\mathrel{\hbox{\rlap{\hbox{\lower4pt\hbox{$\sim$}}}\hbox{$>$}}}}

\newcommand{\mysim}{\mathord{\sim}}

\newcommand{\myapprox}{\mathord{\approx}}
\newcommand{\nimass}{M_\text{Ni56}}
\newcommand{\MWD}{M_{\rm{WD}}}

\title[Sub-Chandrasekhar-mass detonations]{Sub-Chandrasekhar-mass detonations are in tension with the observed $t_0-\nimass$ relation of type Ia supernovae}

\author[D. Kushnir, N. Wygoda and A. Sharon]{
Doron Kushnir$^{1}$\thanks{E-mail: doron.kushnir@weizmann.ac.il}, Nahliel Wygoda$^{2,3}$ and Amir Sharon$^{1}$
\\
$^{1}$Dept. of Particle Phys. \& Astrophys., Weizmann Institute of
Science, Rehovot 76100, Israel\\
$^{2}$Dept. of Astronomy, Yale University, New Haven, CT 06520, USA\\
$^{3}$Dept. of Physics, NRCN, Beer-Sheva 84190, Israel\\
}

\date{Accepted XXX. Received YYY; in original form ZZZ}

\pubyear{2019}

\begin{document}
\label{firstpage}
\pagerange{\pageref{firstpage}--\pageref{lastpage}}
\maketitle

\begin{abstract}
Type Ia supernovae (SNe Ia) are likely the thermonuclear explosions of carbon-oxygen (CO) white-dwarf (WD) stars, but their progenitor systems remain elusive. Recent studies have suggested that a propagating detonation within a thin helium shell surrounding a sub-Chandrasekhar mass CO core can subsequently trigger a detonation within the core (the double-detonation model, DDM). The outcome of this explosion is similar to a central ignition of a sub-Chandrasekhar mass CO WD (SCD). While SCD is consistent with some observational properties of SNe Ia, several computational challenges prohibit a robust comparison to the observations. We focus on the observed $t_0-\nimass$ relation, where $t_0$ (the $\gamma$-rays' escape time from the ejecta) is positively correlated with $\nimass$ (the synthesized $^{56}$Ni mass). We apply our recently developed numerical scheme to calculate SCD and show that the calculated $t_0-\nimass$ relation, which does not require radiation transfer calculations, converges to an accuracy of a few percent. We find a clear tension between our calculations and the observed $t_0-\nimass$ relation. SCD predicts an anti-correlation between $t_0$ and $\nimass$, with $t_0\approx30\,\textrm{day}$ for luminous ($\nimass\gtrsim0.5\,M_{\odot}$) SNe Ia, while the observed $t_0$ is in the range of $35-45\,\textrm{day}$. We show that this tension is larger than the uncertainty of the results, and that it exists in all previous studies of the problem. Our results hint that more complicated models are required, but we argue that DDM is unlikely to resolve the tension with the observations. 
\end{abstract}

\begin{keywords}
hydrodynamics -- shock waves -- supernovae: general 
\end{keywords}



\section{Introduction}
\label{sec:Introduction}

Type Ia supernovae (SNe Ia) are likely the thermonuclear explosions of carbon-oxygen (CO) white-dwarf (WD) stars, but their progenitor systems remain elusive \citep[see][for a review]{Maoz2014}. Sub-Chandrasekhar mass CO WDs have been discussed extensively as a possible progenitor for SNe Ia. Early studies modelled the explosion of sub-Chandrasekhar mass CO WDs with a thick shell of accreted helium and found that a thermonuclear detonation wave (TNDW) in the helium shell can trigger an explosion of the CO core, known as the "double-detonation model" \citep[DDM;][]{Nomoto1982a,Nomoto1982b,Livne1990,Woosley1994}. However, the modelled thick helium shell produces too much $^{56}$Ni during nuclear burning for this to be a viable progenitor \citep{Hoeflich1996,Nugent1997,Kromer2010,Woosley2011}. Recent studies have suggested that the minimal mass of a helium shell required to trigger an explosion in the CO core is much smaller than those used in the early models \citep{Bildsten2007,Fink2007,Fink2010,Moore2013,Shen2014,ShenMoore2014,Polin2019,Townsley2019}, and that only minimal amounts of $^{56}$Ni are synthesized in the helium shell, possibly allowing better agreement with the observations. 

Under the assumption that a TNDW propagating in a very thin shell of helium is sufficient to ignite a second TNDW in the CO core, the outcome of this explosion would be very similar to a central ignition of a sub-Chandrasekhar mass CO WD (sub-Chandra detonation, SCD). One-dimensional (1D) studies of SCD have shown that this model is consistent with some observational properties of SNe Ia, such as the wide range of $^{56}$Ni mass \citep[e.g.,][]{Sim2010,Moll2014,Blondin2017,Shen2018,Bravo2019} and various luminosity-width relations \citep[e.g.,][]{Wygoda2019a,Wygoda2019b}. The simplicity of SCD makes it an ideal benchmark for comparing the results of different numerical codes with each other and with observations. Identifying the observations that are in tension with SCD would be valuable, as this could hint where more complicated models are required. 

Finding an observational quantity that can be robustly compared to a model's predictions is quite challenging. As a demonstration, consider the Phillips relation \citep{Phillips1993}, which relates the maximum flux to the width of the light curve in some band. While this relation can be accurately measured, the prediction of the models is less certain. There are several challenges when it comes to a robust prediction:
\begin{enumerate}
\item The initial profile of the WD is uncertain, as well as some input physical values (e.g., reaction rates) and the ignition location.
\item The calculation of TNDW is challenging \citep[see e.g.,][]{KK2019}, and as a result, it is not clear whether the hydrodynamical calculations converge to the correct values.
\item The radiation transfer calculation is challenging \citep[see e.g., reviews,][]{Hillebrandt00,Noebauer2019}, forcing many uncontrolled approximations, which do not allow a quantitative estimation of the results' uncertainty \citep[for comparisons between various codes, see e.g.,][]{Tanaka2013, Wygoda2019b}. 
\end{enumerate}
For these reasons, the uncertainties involved in a direct comparison of models to the Phillips relation are not well understood. For example,  \citet{Blondin2017} found that SCD models agree well with the Phillips relation for luminous (peak $B$-band magnitude $M_{B}\lesssim-18.5$) SNe Ia but the agreement for dim (peak $M_{B}\gtrsim-18.5$) SNe Ia is not as good (see their figure 5). On the contrary, \citet{Shen2018} found that SCD models agree well with the Phillips relation for dim (peak $M_{B}\gtrsim-19$) SNe Ia but not for luminous (peak $M_{B}\lesssim-19$) SNe Ia (see their figure 14). These conflicting results demonstrate the need for an observational quantity that can be calculated more robustly. 

\begin{figure}
\includegraphics[width=0.48\textwidth]{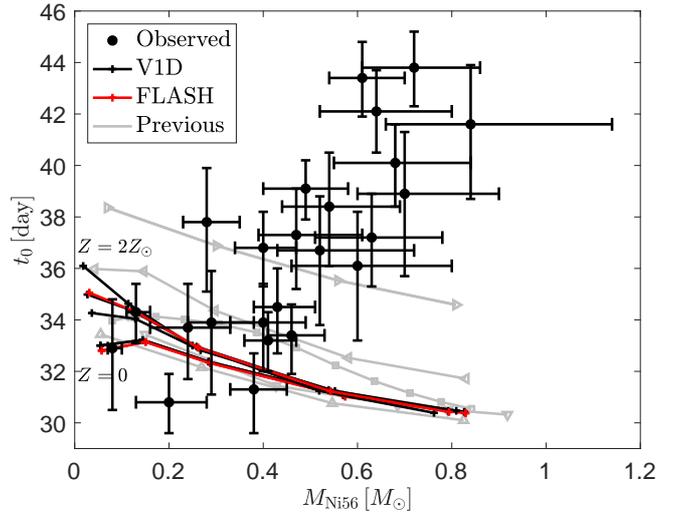}
\caption{The $t_0-\nimass$ relation. Black circles: The observed SNe Ia sample of \citet[][]{Sharon2020}. Plotted are the median of the posterior distribution, together with the $68\%$ confidence levels. Plus signs: The converged SCD results, calculated in V1D (black lines, WD metallicity of $Z=0,0.5,1,2\,Z_{\odot}$) and in FLASH (red lines, WD metallicity of $Z=0,1\,Z_{\odot}$). The metallicity mostly affects the results of the low $\nimass$ cases, where lower $t_0$ values are obtained for lower metallicities. There is a clear tension between the predictions of SCD and the observed $t_0-\nimass$ relation. SCD predicts anti-correlation between $t_0$ and $\nimass$, with $t_0\approx30\,\textrm{day}$ for luminous ($\nimass\gtrsim0.5\,M_{\odot}$) SNe Ia, while the observed $t_0$ is in the range of $35-45\,\textrm{day}$. Grey lines: The results from previous studies of SCD \citep{Sim2010,Moll2014,Blondin2017,Shen2018,Bravo2019} are marked with right-pointing triangles, upward-pointing triangles, squares, left-pointing triangles and downward-pointing triangles, respectively (we thank the authors of these studies for sharing their ejecta profiles with us). The tension with the observed $t_0-\nimass$ relation exists in all previous studies (see Section~\ref{sec:comparison} for detailed discussion).
\label{fig:Nit0}}
\end{figure}

\citet{Stritzinger2006,Scalzo2014,Wygoda2019a} suggested using the $\gamma$-rays (generated in radioactive decays) escape time, $t_0$, defined by \citep{Jeffery1999}
\begin{equation}\label{eq:dep_late}
f_\text{dep}(t) = \frac{t_0^2}{t^2},\;\;\;f_\text{dep}\ll 1,
\end{equation}
where $t$ is the time since explosion and $f_\text{dep}(t)$ is the $\gamma$-ray deposition function, which describes the fraction of the generated $\gamma$-ray energy that is deposited in the ejecta. For a small enough $\gamma$-ray optical depth, each $\gamma$-ray photon has a small chance of colliding with matter from the ejecta (and a negligible chance of additional collisions), such that the deposition function is proportional to the column density, which scales as $t^{-2}$. The value of $t_0$ can be measured from a bolometric light curve to an accuracy of a few percent \citep{Wygoda2019a,Sharon2020} due to an integral relation derived by \citet{Katz2013}, independent of the supernova distance. Together with $\nimass$, the $^{56}$Ni mass synthesized in the explosion \citep[that can be measured to an accuracy of a few tens of percent, e.g.,][]{Sharon2020}, an observed $t_0-\nimass$ relation can be constructed \citep{Wygoda2019a}, see Figure~\ref{fig:Nit0}. The accurate determination of $t_0$ by \citet{Sharon2020} revealed a positive correlation between $t_0$ and $\nimass$. The methods used in previous works did not allow a robust determination of such a correlation, although there were some hints for its existence \footnote{\citet{Stritzinger2006} found a negative correlation between $t_0$ and $\Delta m_{15}(\rm{UVOIR})$ (the decline in pseudo-bolometric magnitude during the first $15$ days after the peak); see their figure 3. \citet{Scalzo2014} found a positive correlation for both $t_0$ and $\nimass$ with SALT2 $x_1$ \citep{Guy2007,Guy2010}, however, the low accuracy of the results diminished the correlation between $t_0$ and $\nimass$ (see their figure 7). The results of \citet{Wygoda2019a} are also not accurate enough to determine the $t_0-\nimass$ correlation (see their figure 5).}. The advantage of comparing models to this observed relation is that it bypasses the need for radiation transfer calculations (challenge (iii)), as the value of $t_0$ can be directly inferred from the ejecta (up to an accuracy of a few percent, see Section~\ref{sec:t0 accuracy}). For example, \citet{Wygoda2019a} showed that Chandrasekhar-mass models deviated significantly from the $t_0-\nimass$ relation for low-luminosity SNe Ia. \citet{Wygoda2019a} also found a small deviation of SCD models from the observed $t_0-\nimass$ relation, but this could not be taken as an evidence for or against SCD models, because of the above-noted (i-ii) challenges. We aim here to resolve challenges (i-ii), in order to allow a robust comparison of the SCD model to the $t_0-\nimass$ relation.  

We have recently developed an accurate and efficient numerical scheme that allows the structure of a TNDW to be resolved \citep{KK2019}. The numerical scheme has two important ingredients: 1. A burning limiter that broadens the width of the TNDW while accurately preserving its internal structure; and 2. An adaptive separation of isotopes into groups that are in nuclear-statistical-quasi-equilibrium (adaptive statistical equilibrium, ASE), which resolves the time-consuming burning calculation of reactions that are nearly balanced-out. The burning limiter limits the changes in both energy and composition to a fraction $f$ during cell sound crossing time (for faster changes, all rates are normalized by a constant factor to limit the changes). Burning is calculated \textit{in situ} by employing the required large-networks without using post-processing or pre-describing the conditions behind the TNDW. In particular, the approach-to and deviation-from nuclear-statistical-equilibrium (NSE) is calculated self-consistently. The scheme was tested against accurate solutions of the structure of a TNDW and against an homologous expansion from NSE, at resolutions typical for multi-dimensional (multi-D) full-star simulations, and an accuracy that is better than a percent for the resolved scales (where the burning limiter is not applied) and a few percent for unresolved scales (broadened by the burning limiter) was obtained. In Section~\ref{sec:setup}, we describe the 1D setup we implement to calculate SCD using two hydrodynamical schemes, VULCAN \citep[Lagrangian, hereafter V1D; for details, see][]{Livne1993IMT} and FLASH4.0 \citep[Eulerian, hereafter FLASH; for details, see][]{Fryxell2000,dubey2009flash}, with the new scheme included. The application of the new scheme resolves the above-noted challenge (ii).  

In Section~\ref{sec:results}, we show that the converging properties (with respect to resolution and the value of $f$) of both codes indicate that the converged results are accurate to better than a few percent. The converged results of these calculations are presented in Figure~\ref{fig:Nit0}, which is the main result of this work. As can be seen in the figure, there is a clear tension between the predictions of SCD and the observed $t_0-\nimass$ relation. SCD predicts an anti-correlation between $t_0$ and $\nimass$, with $t_0\approx30\,\textrm{day}$ for luminous ($\nimass\gtrsim0.5\,M_{\odot}$) SNe Ia, while the observed $t_0$ is in the range of $35-45\,\textrm{day}$

We next show that uncertainties related to challenge (i) are unlikely to resolve the tension with the observations. In Sections~\ref{sec:WD sensitivity}, we study various uncertainties related to the physical processes and to the initial profiles of the WD. We calibrate in Section~\ref{sec:small network} a $69$-isotope network, for which the $t_0-\nimass$ relation is accurately calculated. We then use this reduced network to perform in Section~\ref{sec:reaction rate sensitivity} a sensitivity check of our results to uncertainties in the reaction rate values. We find that the tension between the predictions of SCD and the observed $t_0-\nimass$ relation is much larger than the uncertainty related to the reaction rates.  

In Section~\ref{sec:comparison}, we compare our results to previous studies of the problem performed with less accurate numerical schemes. We show that the general $t_0-\nimass$ (see Figure~\ref{fig:Nit0}) and $\nimass-\MWD$ relations, where $\MWD$ is the mass of the WD, are reproduced in all previous works \citep[except for the results of][which are systematically different from all other works, see Section~\ref{sec:Sim 2010}]{Sim2010}. Specifically, the tension with the observed $t_0-\nimass$ relation exists in all previous studies. The differences between previous works and our results are discussed in detail. We summarise our results in Section~\ref{sec:discussion}, where we argue that the more complicated DDM model is unlikely to resolve the tension with the $t_0-\nimass$ relation. 

In what follows we normalize temperatures, $T_9=T\,[\,\textrm{K}]/10^{9}$, and densities, $\rho_7=\rho\,[\,\textrm{g}\,\textrm{cm}^{-3}]/10^{7}$. Some aspects of this work were calculated with a modified version of the {\sc MESA} code\footnote{Version r7624; https://sourceforge.net/projects/mesa/files/releases/} \citep{Paxton2011,Paxton2013,Paxton2015}. All ejecta profiles used to derive the results in this paper (except for the results in Section~\ref{sec:reaction rate sensitivity}), as well as the bolometric light curves from Section~\ref{sec:t0 accuracy}, are publicly available through \url{https://www.dropbox.com/s/3kd8te2yimdxotm/CIWD.tar.gz?dl=0}. 
 

\section{Numerical schemes and setup}
\label{sec:setup}

In this section, we describe the 1D setup that we implement to calculate SCD using two hydrodynamical schemes. We present our initial setup in Section~\ref{sec:initial} and the ignition method  in Section~\ref{sec:ignition}. The setup of the Lagrangian numerical scheme V1D is described in Section~\ref{sec:V1D} and the setup of the Eulerian numerical scheme FLASH is described in Section~\ref{sec:FLASH}. 

\subsection{Initial setup}
\label{sec:initial}

The WD profiles are constructed using a modified version of a routine by Frank Timmes\footnote{http://cococubed.asu.edu/} that includes the input physics of Appendix~\ref{sec:input}. The WD are isothermal with an initial temperature of $T_{\textrm{WD},9}=0.01$ (the choice of this temperature is discussed in Section~\ref{sec:TWD}, where we also test a different temperature). The initial composition is uniform throughout the WD. We assume that at the time of ignition, the WD contains mainly $^{12}$C and $^{16}$O (typically with equal mass fractions) and some traces of heavier elements, which correspond to the metallicity of the main-sequence progenitor star. Our prescription to determine the initial abundances of the heavy elements is described below, and we show in Section~\ref{sec:initial heavy} that, for our purposes, a few other prescriptions are equivalent, if compared at the same $Y_e$. Following \citet{Timmes2003}, we assume that all the nuclei of $^{12}$C, $^{14}$N, and $^{16}$O present prior to the main-sequence burning are converted to $^{22}$Ne in the WD:
\begin{eqnarray}\label{eq:Ne22}
X\left(^{22}\textrm{Ne}\right)&\approx&22\left[\frac{X_0\left(^{12}\textrm{C}\right)}{12}+\frac{X_0\left(^{14}\textrm{N}\right)}{14}+\frac{X_0\left(^{16}\textrm{O}\right)}{16}\right]\nonumber\\
&\approx&22\left[\frac{X_0\left(\textrm{C}\right)}{12}+\frac{X_0\left(\textrm{N}\right)}{14}+\frac{X_0\left(\textrm{O}\right)}{16}\right],
\end{eqnarray}
where the last approximate equality is because the abundances of the other C,N,O stable isotopes are small \citep{Lodders2003}. For the present-day solar photosphere, we have $X_0(\textrm{C})\approx2.36\times10^{-3}$, $X_0(\textrm{N})\approx6.92\times10^{-4}$, $X_0(\textrm{O})\approx5.73\times10^{-3}$ \citep{Asplund2009}, such that we get $X(^{22}\textrm{Ne})\approx0.0133$ from Equation~\eqref{eq:Ne22}. The solar bulk abundances of the heavy elements are expected to be $\myapprox10\%$ higher than the photospheric values \citep{Turcotte2002}, so in what follows we define 
\begin{eqnarray}\label{eq:Ne22sol}
X\left(^{22}\textrm{Ne}\right)=0.015\left(\frac{Z}{Z_{\odot}}\right).
\end{eqnarray}
Our default composition includes $X(^{22}\textrm{Ne})$, as given by Equation~\eqref{eq:Ne22sol}, and the rest is $^{12}$C and $^{16}$O, such that 
\begin{eqnarray}\label{eq:Yesol}
Y_e=\frac{10}{22}X\left(^{22}\textrm{Ne}\right)+\frac{1}{2}\left[1-X\left(^{22}\textrm{Ne}\right)\right]\approx\frac{1}{2}-6.82\times10^{-4}\left(\frac{Z}{Z_{\odot}}\right).
\end{eqnarray}

\subsection{Ignition method}
\label{sec:ignition}

We ignite a TNDW at the center of the WD by imposing a velocity gradient. This method is different from imposing a temperature hotspot \citep[e.g.,][]{Seitenzahl2009i} at the center of the WD. We find that the velocity method allows smaller ignition regions (with the same resolution and burning limiter), such that even at low resolution the ignition details affecting only a small fraction of the mass. The ignition of a TNDW at the center of a low-resolution WD with a small temperature hotspot was achieved in previous works because a burning limiter was not included \citep[e.g.,][were able to ignite with a $150\,\textrm{km}$ hotspot a $4\,\textrm{km}$ resolution WD]{Miles2019}. In such cases, the ignition is achieved due to a numerical instability, which is suppressed with the burning limiter \citep{Kushnir2013}. The initial velocity profile that we use is linear in the range $[0,r_{\rm{ign}}]$, with $v(0)=0$ and $v(r_{\rm{ign}})=2\times10^{4}\,\textrm{km}\,\textrm{s}^{-1}$. The initial velocity is zero for $r>r_{\rm{ign}}$. In order to suppress the TNDW that propagates to the center of the WD following ignition, we impose a temperature of $4\times10^{9}\,\textrm{K}$ at $r\le r_{\rm{ign}}$ with the composition determined by the NSE conditions. With this choice, the hot material has a small amount of available thermonuclear energy, and the inward propagating TNDW is somewhat suppressed and does not reduce drastically the time-step as it converges to the center. 

We choose for each resolution and burning limiter some small $r_{\rm{ign}}$ that allows ignition. This is done by calibrating for each $\MWD$ a minimal $r_{\rm{ign}}$ in some low-resolution run with $f=0.1$. The calibrated values are given in Table~\ref{tbl:rign}, where the initial cell size in V1D, $\Delta x_{0}$, is defined in Section~\ref{sec:V1D} and the minimal cell size in FLASH, $\Delta x$, is defined in Section~\ref{sec:FLASH}. We scale $r_{\rm{ign}}\propto \Delta x_0/f$ ($r_{\rm{ign}}\propto \Delta x/f$) for different resolutions and $f$ values. This scaling allows us to decrease the ignition region as we increase the resolution, such that the ignition details affect a negligible amount of mass in our converged simulations. We did not decrease $r_{\rm{ign}}$ below some minimal value, given in Table~\ref{tbl:rign}, which is either the critical value for ignition (at any resolution with $f=0.1$) or the minimal value at the converged resolution. 

\begin{table*}
\caption{The calibrated $r_{\rm{ign}}$ values (3rd column) for $f=0.1$ as a function of $\MWD$ (1st column) and the V1D initial (FLASH minimal) cell size, $\Delta x_{0}$ ($\Delta x$), given in the 2nd column. We scale $r_{\rm{ign}}\propto \Delta x_0/f$ ($r_{\rm{ign}}\propto \Delta x/f$) for different resolutions and $f$ values. This scaling allows us to decrease the ignition region as we increase the resolution, such that the ignition details affect but a negligible amount of mass in our converged simulations. We did not decrease $r_{\rm{ign}}$ below some minimal value, given in the 4th column, which is either the critical value for ignition (at any resolution with $f=0.1$) or the minimal value at the converged resolution.}
\begin{tabular}{|c||c||c||c|}
\hline
$\MWD\,[M_{\odot}]$  &    Resolution [km]    & $r_{\rm{ign}}$ for $f=0.1$ [km] & Minimal $r_{\rm{ign}}$ [km] \\ \hline
0.8	&  $\Delta x_{0}=6.98$   &	200 & 12.5	\\
        &  $\Delta x=4$   &	200 & 12.5	\\
0.85	&  $\Delta x_{0}=6.59$   &	200 & 6.25	\\
        &  $\Delta x=4$   &	200 & 6.25	\\
0.9	&  $\Delta x_{0}=6.21$   &	200 & 6.25	\\
        &  $\Delta x=4$   &	200 & 6.25	\\
1	&  $\Delta x_{0}=5.45$   &	100 & 12.5	\\
        &  $\Delta x=4$   &	100 & 12.5	\\
1.1	&  $\Delta x_{0}=4.67$   &	100 & 6.25	\\
        &  $\Delta x=4$   &	100 & 6.25	\\
\hline
\end{tabular}
\centering
\label{tbl:rign}
\end{table*}

\subsection{Lagrangian code -- VULCAN}
\label{sec:V1D}

We use our modified V1D version \citep{KK2019} that is compatible with the input physics of Appendix~\ref{sec:input}. Our default isotope list is the NSE$5$ list of $179$ isotopes \citep{Kushnir2019} without $^{6}$He ($178$ isotopes in total). Unless stated otherwise, we ignore weak reactions and thermal neutrino emission (we show in Section~\ref{sec:uncertainty} that it is safe to ignore these effects). We do not use linear artificial viscosity, the Courant time-step factor is $0.25$, and the maximum relative change of the density in each cell during a time-step is set to $0.01$. Burning is not allowed on shocks (identified as cells where $q_{v}/p>0.1$, where $q_{v}$ is the artificial viscosity and $p$ is the pressure). The allowed error tolerance for the burning integration is $\delta_{B}=10^{-8}$ \citep[see][for details]{KK2019}. 

The mesh includes only the WD, with the outer numerical node at the surface of the WD. The inner boundary condition is of a solid wall and the outer boundary condition is of a free surface. Initially, all cells are of equal size, $\Delta x_{0}$, and the density in each cell is determined by interpolation from the original WD profile to the center of the cell. We then redefine the mesh for cells with $\rho_7<0.01$, such that these cells have the same mass, which is equal to the mass of the outermost cell with $\rho_7\ge0.01$. The radii of these cells are determined by interpolation of the original WD profile. This allows us to significantly increase the size of the outer cells (and increase the time-step when the shock propagates through these cells) without decreasing the mass resolution. 

Since the initial profile is interpolated to the mesh, it is not in strict hydrostatic equilibrium. We therefore only activate cells that are just in-front of the leading shock. This is done by finding the outermost active cell with $q_{v}/p>10^{-3}$ and then activating its outer node. Initially, all cells within $[0,r_{\rm{ign}}+5\Delta x_{0}]$ are activated. 

We examine throughout the simulation the total kinetic energy, $E_{\rm{kin}}$, the total internal energy, $E_{\rm{int}}$, and the total gravitational energy, $E_{\rm{grav}}$. We stop the simulation when both $E_{\rm{kin}}/E_{\rm{int}}>20$ and $-E_{\rm{kin}}/E_{\rm{grav}}>20$ (typically the former condition is fulfilled later). At this point, the deviations from homologous expansion are of a few percent. The velocity of each node, $v_i$, for the asymptotic freely expanding ejecta is determined by $v_i=r_i/t_{\rm{eff}}$, where $r_i$ is the radius of each node and $t_{\rm{eff}}$ is determined such that the total kinetic energy of the asymptotic ejecta equals $E_{\rm{kin}}$. 

\subsection{Eulerian code -- FLASH}
\label{sec:FLASH}

We use our modified FLASH version \citep{KK2019} that is compatible with the input physics of Appendix~\ref{sec:input}. Specifically, instead of using the supplied burning routines of FLASH, which only support hard-wired $\alpha$-nets, we use the burning routines of V1D with the same integration method. 

The simulations are performed in spherical geometry, the cutoff value for the composition mass fraction is $\textsc{smallx}=10^{-25}$, and the Courant time-step factor is $\textsc{CFL}=0.2$. Burning is not allowed on shocks and the nuclear burning time-step factor is $\textsc{enucDtFactor}=0.2$. 

The computed region is between $x=0$ and $x=2^{17}\,\textrm{km}\approx1.31\times10^{5}\,\textrm{km}$. The WD profile is interpolated into the mesh, and the region outside the WD has $\rho_7=10^{-11}$ and $T_{9}=0.01$. We use $16$ cells per block and a minimal refinement level of $8$, such that the minimal resolution is $64\,\textrm{km}$. The maximal resolution, $\Delta x$, is determined by the maximal refinement level, and can be reached according to the refinement conditions. In order to determine whether we define or redefine a block, we go over all the cells within the block and find the minimal radius, $x_{\min}$, the maximal density, $\rho_{\max}$, the minimal burning limiter, $f_{lim}$, and the minimal burning limiter calculated with a factor of two coarser resolution, $f_{lim,2}$. We then use the following scheme for the refinement (each condition supersedes all previous conditions):
\begin{enumerate}
\item A density gradient refinement condition with $\textsc{refine\_cutoff}=0.8$, $\textsc{derefine\_cutoff}=0.2$, and $\textsc{refine\_filter}=0.01$.
\item If $x_{\min}\le r_{\rm{ign}}$ and $t<0.1\,\textrm{s}$, refine. This is done to ensure the highest resolution in the region of ignition. 
\item If $\rho_{\max,7}<0.01$, derefine.
\item If $f_{lim}<0.98$, refine.
\item If $f_{lim,2}<0.98$, do not derefine.
\end{enumerate}
Our scheme ensures that whenever the burning limiter is active (i.e., the relevant parameters are changing faster than the sound crossing time), the resolution is maximal. We decrease the resolution in regions with low ($\rho_7<0.01$) density. We show in Section~\ref{sec:uncertainty} that the observables of interest are accurately calculated with this refinement scheme. 

 The inner boundary condition is "reflected" (a solid wall), and the outer boundary condition is that of a free flow ("diode"). Since the initial profile is interpolated to the mesh, it is not in strict hydrostatic equilibrium. We therefore override in each time-step any deviations from the initial conditions of un-shocked cells. In this way, cells always have the initial upstream conditions up to the point where the shock crosses them. This can be enforced up to the time when the shock is a few cells away from the WD surface. We stop the simulation when both $E_{\rm{kin}}/E_{\rm{int}}>20$ and $-E_{\rm{kin}}/E_{\rm{grav}}>20$ (typically the former condition is fulfilled later). This condition is reached when less than $0.1\%$ of the mass has left the computed region. We define the velocity of each node $v_i=r_i/t_{\rm{eff}}$ for the asymptotic freely expanding ejecta, similarly to the V1D case.
 

\section{The predicted $\lowercase{t}_0-\nimass$ relation}
\label{sec:results}

In this section, we present our results for the $t_0-\nimass$ relation of SCD. In Section~\ref{sec:convergence}, we study the converging properties of the simulations, and we show that our results are converged to a few percent. The converged results are presented in Section~\ref{sec:default}. Several different ways to estimate $t_0$ are compared in Section~\ref{sec:t0 accuracy}, and we show that all of them are in agreement with the level of a few percent. 

\subsection{Convergence study}
\label{sec:convergence}

We calculate for five WD masses, $\MWD=0.8,0.85,0.9,1,1.1\,M_{\odot}$, and two metallicities, $Z=0,1Z_{\odot}$ ($10$ cases in total), with our default input physics. For each case, we use both V1D and FLASH with $f=0.1$, and with different resolutions. From the asymptotic freely expanding ejecta of each calculation, we determine $\nimass$\footnote{Note that since weak reactions are not included, a small amount of mass is located in $^{56}$Cu with a half life of $93\,\textrm{ms}$. We therefore add this mass to $\nimass$.} and $t_0$, which is given by \citep{Wygoda2019a}:
\begin{eqnarray}\label{eq:sigmav}
t_0^2=\frac{\kappa_{\rm{eff}}}{\nimass}\int_{0}^{\infty} dv v^2\rho(v)X_{\rm{Ni}56}\left(v\right)\int d\hat{\Omega}\int_{0}^{\infty}ds\rho(\vec{v}+s\hat{\Omega}),
\end{eqnarray}
with $\kappa_{\rm{eff}}\approx0.025(Y_e/0.5)\,\textrm{cm}^2\,\textrm{g}^{-1}$ \citep[][we use $Y_{e}=0.5$ when evaluating Equation~\eqref{eq:sigmav} in what follows]{Swartz1995,Jeffery1999}. Our lowest resolution calculations have $\Delta x_0=R_{\rm{WD}}/1000$ (where $R_{\rm{WD}}$ is the initial radius of the WD, V1D) or $\Delta x=4\,\textrm{km}$ (FLASH). We perform higher resolution calculations, increasing the resolution by a factor of two each time, until we reach convergence in $\nimass$ (the convergence of $t_0$ is faster) to a level better than a few percent. In most cases, the convergence is on the sub-percent level. We then repeat all the calculations with $f=0.05$ and with $f=0.025$ (for the same resolutions). 

\begin{figure}
\includegraphics[width=0.48\textwidth]{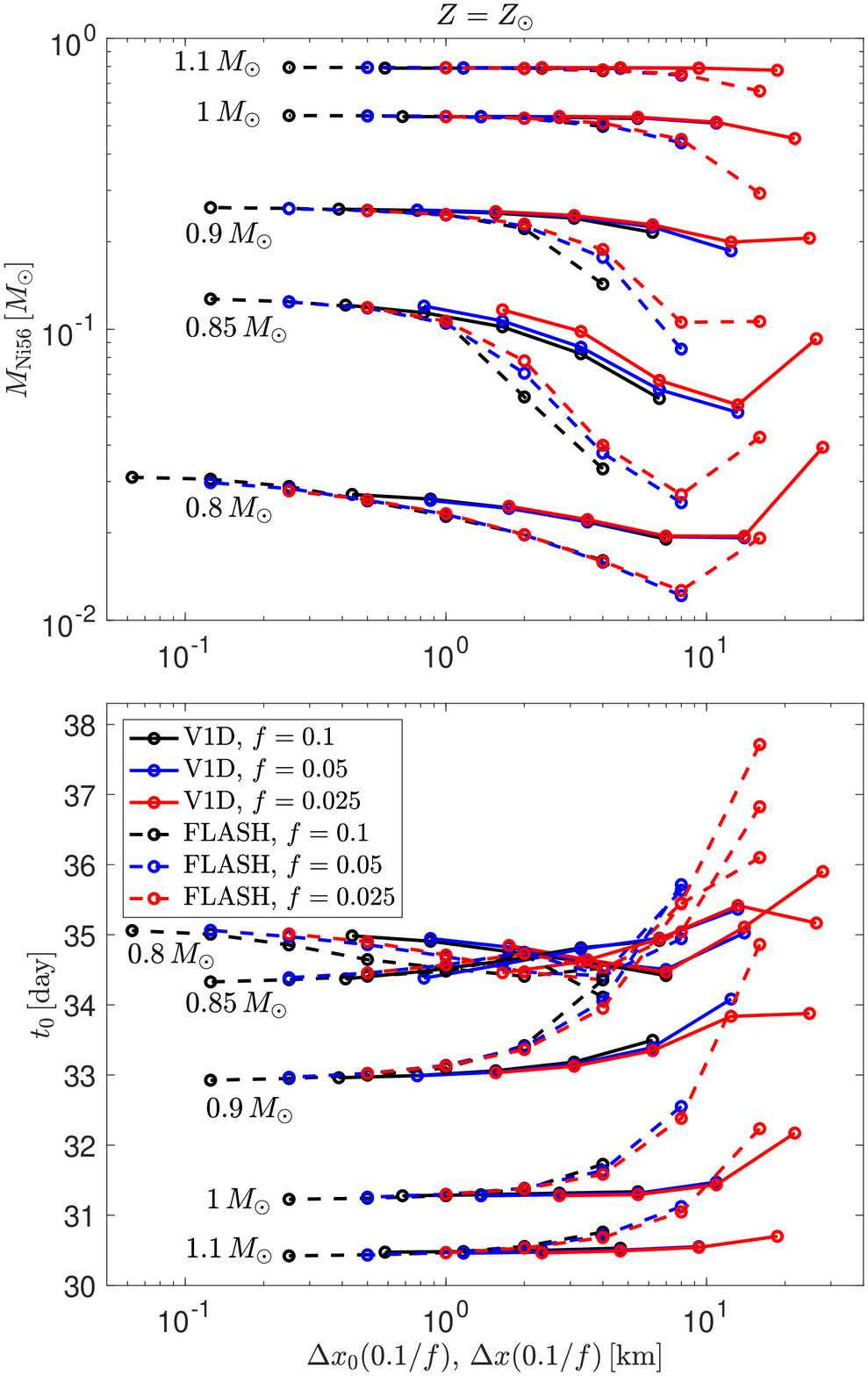}
\caption{$\nimass$ (top panel) and $t_0$ (bottom panel) convergence tests for $Z=Z_{\odot}$. Since the burning limiter uses $\mysim 1/f$ cells to describe the fast burning region, it is expected that the resolving power of the calculation will decrease linearly with $f$. We therefore plot $\nimass$ and $t_0$ as a function of $\Delta x_{0} (0.1/f)$ for V1D (solid lines) or $\Delta x (0.1/f)$ for FLASH (dashed lines). Different colors (black, blue, red) correspond to different $f$ values ($0.1,0.05,0.025$, respectively). Five WD masses, $\MWD=0.8,0.85,0.9,1,1.1\,M_{\odot}$, are considered. The V1D results converge with $\Delta x_{0}(0.1/f)\sim0.5\,\textrm{km}$ and the FLASH results converge with $\Delta x(0.1/f)\sim0.1\,\textrm{km}$. The difference in the required resolutions between V1D and FLASH corresponds to the $\mysim2-4$ compression factor behind the leading shock of the TNDW. 
\label{fig:Ni-t0_Convergence_Z1}}
\end{figure}

The convergence test for $Z=Z_{\odot}$ is presented in Figure~\ref{fig:Ni-t0_Convergence_Z1}. Since the burning limiter uses $\mysim 1/f$ cells to describe the fast burning region, it is expected that the resolving power of the calculation will decrease linearly with $f$. We therefore plot $\nimass$ and $t_0$ as a function of $\Delta x_{0} (0.1/f)$ (or $\Delta x (0.1/f)$), and indeed the scaled results are roughly $f$ independent. The V1D results converge with $\Delta x_{0}(0.1/f)\sim0.5\,\textrm{km}$ and the FLASH results converge with $\Delta x(0.1/f)\sim0.1\,\textrm{km}$. The difference in the required resolutions between V1D and FLASH corresponds to the $\mysim2-4$ compression factor behind the leading shock of the TNDW. The converged values of $\nimass$ and $t_0$ are presented in Table~\ref{tbl:Ni56 Conv Z1}, along with the required resolutions for convergence to the indicated level (estimated by using the results with a factor-of-two coarser resolution). We provide the results for $f=0.1$, as the results with smaller values of $f$ seem to converge to the same values (but would require higher resolution for convergence). The only exceptions are the V1D calculations of $\MWD=1,1.1\,M_{\odot}$, where there is a subtle, sub-percent, difference between the $f=0.1$ and the $f=0.05,0.025$ results, which may be related to the erroneous behaviour of V1D with $f=0.1$ in high densities \citep[see section 4.5 of][]{KK2019}. We therefore conservatively use $f=0.05$ in these cases. Similar results for the $Z=0$ case are presented in Appendix~\ref{sec:more results} (Figure~\ref{fig:Ni-t0_Convergence_Z0} and Table~\ref{tbl:Ni56 Conv Z0}).

\begin{table*}
\begin{minipage}{160mm}
\caption{The converged $Z=Z_{\odot}$ values of $\nimass$ (4th column) and $t_0$ (5th column), along with the convergence level (in parenthesis, estimated as the deviation from the results with a factor-of-two coarser resolution), as a function of $\MWD$ (1st column). The required V1D initial (FLASH minimal) cell sizes, $\Delta x_{0}$ ($\Delta x$), for convergence are given in the 2nd column. The $f$ values of the converged calculations are given in the 3rd column (see text). The $t_0$ values estimated with the MC $\gamma$-ray transport (full radiation transfer) calculations are given in the 6th (7th) column.}
\begin{tabular}{|c||c||c||c||c||c||c|}
\hline
$\MWD\,[M_{\odot}]$  &$\Delta x_0,\,\Delta x\,[\textrm{km}]$ & $f$ & $\nimass\,[M_{\odot}]$ & $t_0\,[\textrm{day}]$ & $t_0^{\gamma\rm{RT}}\,[\textrm{day}]$ & $t_0^{\rm{RT}}\,[\textrm{day}]$ \\ \hline
0.8	&	$\Delta x_0$: 0.44	&	0.1	&	0.027	&	35.0	&	34.9	&	34	\\
	&		&		&	(3.68\%)	&	(0.22\%)	&		&		\\
	&	$\Delta x$: 0.0625	&	0.1	&	0.031	&	35.1	&	34.7	&	34	\\
	&		&		&	(1.49\%)	&	(0.15\%)	&		&		\\
0.85	&	$\Delta x_0$: 0.41	&	0.1	&	0.121	&	34.4	&	34.1	&	34	\\
	&		&		&	(6.05\%)	&	(0.32\%)	&		&		\\
	&	$\Delta x$: 0.125	&	0.1	&	0.127	&	34.3	&	33.8	&	34	\\
	&		&		&	(2.16\%)	&	(0.09\%)	&		&		\\
0.9	&	$\Delta x_0$: 0.39	&	0.1	&	0.259	&	33.0	&	32.5	&	32	\\
	&		&		&	(0.96\%)	&	(0.09\%)	&		&		\\
	&	$\Delta x$: 0.125	&	0.1	&	0.262	&	32.9	&	32.4	&	32	\\
	&		&		&	(0.68\%)	&	(0.07\%)	&		&		\\
1	&	$\Delta x_0$: 0.68	&	0.05	&	0.539	&	31.3	&	30.6	&	31	\\
	&		&		&	(0.30\%)	&	(0.05\%)	&		&		\\
	&	$\Delta x$: 0.25	&	0.1	&	0.543	&	31.2	&	30.8	&	31	\\
	&		&		&	(0.18\%)	&	(0.05\%)	&		&		\\
1.1	&	$\Delta x_0$: 0.58	&	0.05	&	0.792	&	30.5	&	29.7	&	31	\\
	&		&		&	(0.06\%)	&	(0.05\%)	&		&		\\
	&	$\Delta x$: 0.25	&	0.1	&	0.794	&	30.4	&	29.8	&	31	\\
	&		&		&	(0.09\%)	&	(0.05\%)	&		&		\\
\hline
\end{tabular}
\centering
\label{tbl:Ni56 Conv Z1}
\end{minipage}
\end{table*}

The difference between the V1D- and the FLASH-converged results is usually consistent with the level of convergence. The exception is the $M=0.8\,M_{\odot}$ case, where there is a $\myapprox15\%$ deviation in $\nimass$ between the two codes, which is a factor of few larger than the convergence level estimate of each result. We believe that the V1D result is more accurate because of the high accuracy of energy conservation ($\mysim10^{-6}$) obtained in these calculations, as compared to the FLASH calculations ($\mysim10^{-3}$). When recalculating the FLASH sequence with $\textsc{CFL}=0.1$ (the default calculations are with $\textsc{CFL}=0.2$), the deviation between the two codes is only slightly reduced. We perform more numerical tests in Section~\ref{sec:uncertainty}, but we are unable to locate the exact reason for the deviation between the two codes in this case. Nevertheless, none of our conclusions is sensitive to this deviation.  

The burning limiter guarantees that, as long as the small burning scale (i.e., where the burning limiter is operating) is in steady state (meaning that the solution in this region does not change while the region propagates to a few times its own size), the solution is accurate (or at least converges very fast to the correct solution), since the solution is independent of the reaction rates \citep{KK2019}. We would therefore expect that the solution will converge to $\mysim1\%$ when the WD is resolved with $\mysim100/f$ cells, such that $\Delta x\sim fR_{\rm{WD}}/100\sim(f/0.1)5\,\textrm{km}$. The information presented in Table~\ref{tbl:Ni56 Conv Z1} suggests that this naive expectation is (only) a factor of few lower than the convergence properties of $t_0$ for all WD masses and $\nimass$ for high WD masses ($\mysim0.1\%$ convergence for $\Delta x\mysim0.1\,\textrm{km}$). However, much a higher resolution is required for the $\nimass$ convergence of low WD masses. The reason for this higher resolution requirement is related to the $^{56}$Ni mass distribution
within the ejecta, and is explained below.

\begin{figure}
\includegraphics[width=0.48\textwidth]{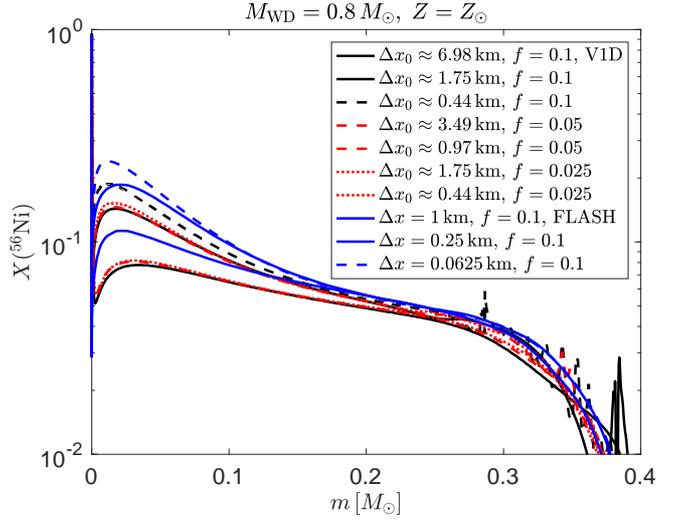}
\caption{The $^{56}$Ni mass fraction distribution, $X(^{56}\rm{Ni})$, within the $\MWD=0.8\,M_{\odot}$, $Z=Z_{\odot}$ ejecta, as a function of the mass coordinate, $m$. Plotted are V1D results with $f=0.1,0.05,0.025$ values (black, dashed-red and dotted-red lines, respectively) and with different initial resolutions. The results follow the scaling $\Delta x_{0}/f$ (compare the red-dashed and -dotted lines to the black lines). The region around $m=0.2\,M_{\odot}$ converges rapidly, since in this region the small burning scale is very close to steady state. At smaller $m$, the small burning scale is further away from steady state, so higher resolution is required for convergence. For $m\gtrsim0.3\,M_{\odot}$, the $X(^{56}\rm{Ni})$ distribution seems irregular for some V1D calculations. This is because the TNDW becomes unstable at low upstream densities, just before it dies out. The behaviour of the FLASH calculations (blue lines) is similar to that of the V1D calculations, although they converge to a slightly higher $X(^{56}\rm{Ni})$; see discussion in the text. 
\label{fig:EjectaConvergence_M08_Z1}}
\end{figure}

The convergence properties of our calculations are further studied in Figure~\ref{fig:EjectaConvergence_M08_Z1}, in which the $^{56}$Ni mass fraction distribution, $X(^{56}\rm{Ni})$, within the $\MWD=0.8\,M_{\odot}$ ejecta, as a function of the mass coordinate, $m$, is presented. As can be seen in the figure, the V1D results with different $f$ values follow the scaling $\Delta x_{0}/f$ (compare the red-dashed and -dotted lines to the black lines). The region around $m=0.2\,M_{\odot}$ converges rapidly, since in this region the small burning scale is very close to steady state. At smaller $m$, the small burning scale is further away from steady state, so higher resolution is required for convergence. As we approach the ignition region, increasingly higher resolution is required for ignition, up to the innermost region, where the steady-state assumption completely fails, and a resolution comparable to the burning scale of a TNDW ($\mysim1\,\textrm{cm}$) is required. Nevertheless, the mass within $r_{\rm{ign}}$ is $<2\times10^{-4}\,M_{\odot}$, and so the mass that is not resolved correctly with the presented resolution is negligible, and the integral properties of the $^{56}$Ni mass distribution converge fast to the correct values. Since a large fraction of the $^{56}$Ni mass is within a region that is not in a strict steady state, higher resolution than the naive expectation above is required for convergence. 

For $m\gtrsim0.3\,M_{\odot}$, the $X(^{56}\rm{Ni})$ distribution seems irregular for some V1D calculations. This is because the TNDW becomes unstable at low upstream densities, just before it dies out \citep[see also][for instability at high upstream densities]{Khokhlov93}. We seem to capture this process with the V1D calculations, but since it is quite random, a convergence study in this region is more problematic. Nevertheless, the integral properties of the $^{56}$Ni mass distribution are hardly affected by the exact process in which the TNDW dies out. It should also be noted that there are transverse modes of instability for TNDW \citep{Boisseau96,Gamezo99,Timmes2000} that are not captured in our 1D calculations. The behaviour of the FLASH calculations (blue lines) is similar to the V1D calculations, although they converge to a slightly higher $X(^{56}\rm{Ni})$, as discussed above. 

\begin{figure}
\includegraphics[width=0.48\textwidth]{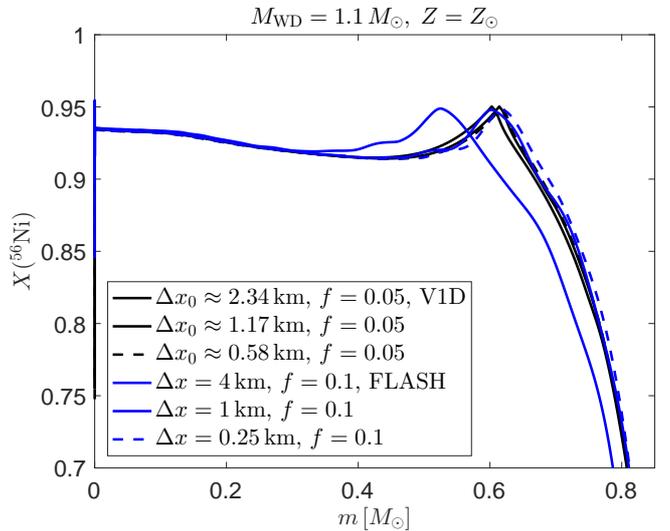}
\caption{The $^{56}$Ni mass fraction distribution, $X(^{56}\rm{Ni})$, within the $\MWD=1.1\,M_{\odot}$, $Z=Z_{\odot}$ ejecta, as a function of the mass coordinate, $m$. V1D (FLASH) results with $f=0.05(0.1)$ and different initial (maximal) resolutions are plotted in black (blue) lines. $X(^{56}\rm{Ni})$ almost reaches unity for a large fraction of the mass, and the convergence is fast. Both V1D and FLASH converge to the same values. 
\label{fig:EjectaConvergence_M11_Z1}}
\end{figure}

We provide another example in Figure~\ref{fig:EjectaConvergence_M11_Z1}, in which a similar convergence study for the $\MWD=1.1\,M_{\odot}$ case is presented. In this case, $X(^{56}\rm{Ni})$ almost reaches unity for a large fraction of the mass, and it seems that the steady-state assumption is accurate for the majority of the $^{56}$Ni mass. Therefore, the convergence in this case is faster and agrees with the naive expectation noted above. Both V1D and FLASH converge to the same values. 

\subsection{Results for the default setup}
\label{sec:default}

\begin{figure}
\includegraphics[width=0.48\textwidth]{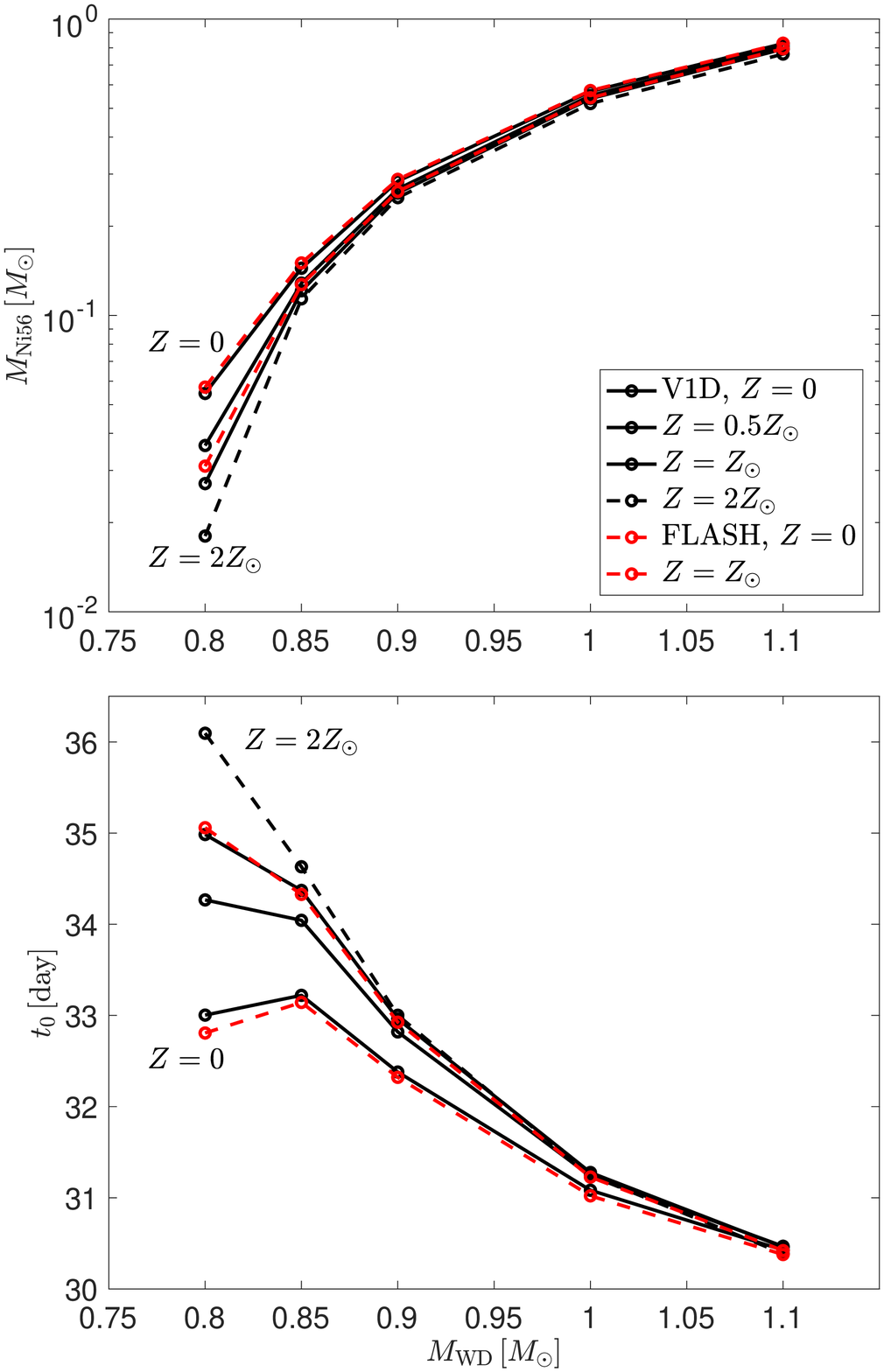}
\caption{The converged SCD $\nimass$ and $t_0$ values (top and bottom panel, respectively), calculated in V1D (black lines, WD metallicity of $Z=0,0.5,1,2\,Z_{\odot}$) and in FLASH (red lines, WD metallicity of $Z=0,1\,Z_{\odot}$), as a function of $\MWD$. The metallicity mostly affects the results of the low $\nimass$ cases, where higher $\nimass$ (lower $t_0$) values are obtained for lower metallicities. $\nimass$ is a strong function of $M_{\rm{WD}}$, while $t_0$ only changes by $\mysim20\%$.
\label{fig:ConvergedNit0M}}
\end{figure}

We calculate with V1D two more metallicities ($Z=0.5Z_{\odot}$ and $Z=2Z_{\odot}$) for the five WD masses. For each case, we use the required resolutions and $f$ values for convergence as determined from the  $Z=0,Z_{\odot}$ cases (see Tables~\ref{tbl:Ni56 Conv Z1} and~\ref{tbl:Ni56 Conv Z0}), and the results are given in Tables~\ref{tbl:Ni56 Conv Z05} and~\ref{tbl:Ni56 Conv Z2}. The converged results of all cases are presented in Figures~\ref{fig:Nit0} and~\ref{fig:ConvergedNit0M}. As can be seen in Figure~\ref{fig:ConvergedNit0M}, $\nimass$ is a strong function of $M_{\rm{WD}}$, while $t_0$ only changes by $\mysim20\%$. The metallicity mostly affects the results of the low $\nimass$ cases. In Figure~\ref{fig:Nit0}, the $t_0-\nimass$ relation is compared to the observed sample of \citet{Sharon2020}. As can be seen in the figure, there is a clear tension between the predictions of SCD and the observed $t_0-\nimass$ relation. SCD predicts an anti-correlation between $t_0$ and $\nimass$, with $t_0\approx30\,\textrm{day}$ for luminous ($\nimass\gtrsim0.5\,M_{\odot}$) SNe Ia, while the observed $t_0$ is in the range of $35-45\,\textrm{day}$. In the following sections, we show that this tension is larger than the uncertainty of the results. 

\subsection{The accuracy of inferring $t_0$ using Equation~\eqref{eq:sigmav}}
\label{sec:t0 accuracy}

One caveat of the comparison in Figure~\ref{fig:Nit0} is that the calculated $t_0$ are estimated with Equation~\eqref{eq:sigmav}, while the observed $t_0$ are extracted from the bolometric light curves. In order to estimate the uncertainty associated with this, we perform Monte-Carlo (MC) $\gamma$-ray transport calculations to determine $f_{\rm{dep}}$ for the converged asymptotic freely expanding ejecta, using the methods described in \citet{Sharon2020}. At late times, we get $f_{\rm{dep}}\propto t^{-2}$, so we determine $t_{0}^{\gamma\rm{RT}}=f_{\rm{dep}}^{1/2}t$. The obtained values are presented in Tables~\ref{tbl:Ni56 Conv Z1}, \ref{tbl:Ni56 Conv Z0}, \ref{tbl:Ni56 Conv Z05}, and~\ref{tbl:Ni56 Conv Z2}. In all cases, the deviation of $t_{0}^{\gamma\rm{RT}}$ from $t_0$ is smaller than $3\%$, where the largest deviations are for high-luminosity SNe Ia with $t_{0}^{\gamma\rm{RT}}$ systematically smaller than $t_0$ (increasing the tension with observations). The main reason for this deviation is the approximation $Y_{e}=0.5$ used in Equation~\eqref{eq:sigmav}, where for high $\nimass$, a significant fraction of the ejecta has decayed to $^{56}$Fe with $Y_{e}\approx0.46$, such that $t_0$ can be smaller by up to $\myapprox4\%$. We find from the MC simulations that both the photoelectric effect and relativistic corrections, of order $v/c$, slightly increase the $t_0$. 

We next use the full radiation transfer code URILIGHT \citep{Wygoda2019a} to calculate bolometric light curves for our ejecta (see Appendix~\ref{sec:full RT} for details regarding the radiation transfer calculations). The uncertainties associated with the  full radiation transfer calculation are hard to estimate, which is the reason we bypass this step with the calculation of $t_0$. Nevertheless, we use here the full radiation transfer calculation as a sanity check. We use the same methods of \citet{Sharon2020} to extract the $\gamma$-ray deposition history from the bolometric light curve. The $\gamma$-ray escape time from this procedure, $t_{0}^{\rm{RT}}$, is only accurate to about $1\,\textrm{day}$ (on top of the uncertainties related to the calculation of the light curve, see detailed discussion in Appendix~\ref{sec:full RT}). The obtained results, presented in Tables~\ref{tbl:Ni56 Conv Z1}, \ref{tbl:Ni56 Conv Z0}, \ref{tbl:Ni56 Conv Z05}, and~\ref{tbl:Ni56 Conv Z2}, are consistent with both $t_0$ and $t_{0}^{\gamma\rm{RT}}$. 

We conclude that Equation~\eqref{eq:sigmav} is accurate to a few percent, with values that are systematically higher than $t_{0}^{\gamma\rm{RT}}$, such that using Equation~\eqref{eq:sigmav} only decreases the tension with the observations.


\section{The uncertainty of the results}
\label{sec:WD sensitivity}

In this section, we estimate the uncertainty of our results due to various effects. We study the sensitivity to several numerical and physical processes in Section~\ref{sec:uncertainty} and to the initial heavy element abundances in Section~\ref{sec:initial heavy}. Larger changes of the initial WD profile are studied in Section~\ref{sec:initial profile}. 

\subsection{Sensitivity to numerical and physical parameters}
\label{sec:uncertainty}

The numerical calculations include some numerical and physical parameters that are uncertain. We choose for each $\MWD$ and metallicity ($Z=0$ and $Z=Z_{\odot}$) a V1D initial resolution for which the deviation of the results ($\nimass$ and $t_0$) from the converged values is $\lesssim5\%$. For each case, we calculate with V1D the sensitivity of the results to a few uncertainties (when relevant, the initial structure of the WD changes as well). The $Z=Z_{\odot}$ results for $\nimass$ and $t_0$ are presented in Tables~\ref{tbl:Ni56 Sens Z1} and~\ref{tbl:t0 Sens Z1}, respectively (similar results are presented for $Z=0$ in Appendix~\ref{sec:more results}, Tables~\ref{tbl:Ni56 Sens Z0}-\ref{tbl:t0 Sens Z0}). 

\begin{table*}
\begin{minipage}{160mm}
\caption{Sensitivity of the calculated $\nimass$ (in $M_{\odot}$) of the $Z=Z_{\odot}$ case to various numerical and physical parameters. We choose for each $\MWD$ (1st column) a V1D initial resolution (2nd column) for which the deviation of the results from the converged values is $\lesssim5\%$ (3rd column, the deviation in parenthesis). For each case, we calculate with V1D the sensitivity of $\nimass$ to several uncertainties (when relevant, the initial structure of the WD changes as well), ordered in the 4th-11th column (deviation from the reference case in parenthesis) as follows:  isotope list NSE6, isotope list NSE7, without using the ASE scheme, adding weak nuclear reactions, adding thermal neutrino emission, without Coulomb corrections to the EOS, without nuclear excitation energy contribution to the EOS, and without nuclear reaction screening.}
\begin{tabular}{|c||c||c||c||c||c||c||c||c||c||c|}
\hline
$\MWD\,[M_{\odot}]$  & $\Delta x_0\,[\textrm{km}]$ & Reference & NSE6  & NSE7 & w/o ASE & w weak & w thermal $\nu$ & w/o Coul. & w/o ex. & w/o screen \\ \hline
0.8	&	0.87	&	0.0261	&	0.0261	&	0.0261	&	0.0261	&	0.0261	&	0.0261	&	0.0160	&	0.0266	&	0.0233	\\
	&		&	(3.68\%)	&	(0.000\%)	&	(0.000\%)	&	(0.000\%)	&	(0.022\%)	&	(0.002\%)	&	(48.0\%)	&	(1.80\%)	&	(11.5\%)	\\
0.85	&	0.82	&	0.1140	&	0.1140	&	0.1140	&	0.1140	&	0.1140	&	0.1140	&	0.0478	&	0.1162	&	0.0912	\\
	&		&	(6.05\%)	&	(0.003\%)	&	(0.001\%)	&	(0.021\%)	&	(0.028\%)	&	(0.004\%)	&	(81.9\%)	&	(1.87\%)	&	(22.2\%)	\\
0.9	&	1.55	&	0.2509	&	0.2509	&	0.2509	&	0.2509	&	0.2508	&	0.2509	&	0.1741	&	0.2526	&	0.2273	\\
	&		&	(3.09\%)	&	(0.001\%)	&	(0.001\%)	&	(0.004\%)	&	(0.042\%)	&	(0.005\%)	&	(36.2\%)	&	(0.69\%)	&	(9.9\%)	\\
1	&	2.73	&	0.5334	&	0.5334	&	0.5334	&	0.5334	&	0.5331	&	0.5334	&	0.4534	&	0.5345	&	0.5111	\\
	&		&	(1.03\%)	&	(0.000\%)	&	(0.000\%)	&	(0.004\%)	&	(0.055\%)	&	(0.005\%)	&	(16.2\%)	&	(0.21\%)	&	(4.3\%)	\\
1.1	&	2.34	&	0.7912	&	0.7912	&	0.7912	&	0.7911	&	0.7906	&	0.7911	&	0.7334	&	0.7916	&	0.7833	\\
	&		&	(0.16\%)	&	(0.001\%)	&	(0.000\%)	&	(0.008\%)	&	(0.077\%)	&	(0.008\%)	&	(7.6\%)	&	(0.05\%)	&	(1.0\%)	\\
\hline
\end{tabular}
\centering
\label{tbl:Ni56 Sens Z1}
\end{minipage}
\end{table*}

\begin{table*}
\begin{minipage}{160mm}
\caption{Same as Table~\ref{tbl:Ni56 Sens Z1} for $t_0$ (in day).}
\begin{tabular}{|c||c||c||c||c||c||c||c||c||c||c|}
\hline
$\MWD\,[M_{\odot}]$  & $\Delta x_0\,[\textrm{km}]$ & Reference & NSE6  & NSE7 & w/o ASE & w weak & w thermal $\nu$ & w/o Coul. & w/o ex. & w/o screen \\ \hline
0.8	&	0.87	&	34.91	&	34.91	&	34.91	&	34.91	&	34.91	&	34.91	&	35.62	&	34.92	&	35.02	\\
	&		&	(0.22\%)	&	(0.000\%)	&	(0.000\%)	&	(0.001\%)	&	(0.001\%)	&	(0.001\%)	&	(2.0\%)	&	(0.03\%)	&	(0.3\%)	\\
0.85	&	0.82	&	34.48	&	34.48	&	34.48	&	34.48	&	34.48	&	34.48	&	35.73	&	34.45	&	34.89	\\
	&		&	(0.32\%)	&	(0.000\%)	&	(0.000\%)	&	(0.001\%)	&	(0.002\%)	&	(0.002\%)	&	(3.6\%)	&	(0.09\%)	&	(1.2\%)	\\
0.9	&	1.55	&	33.06	&	33.06	&	33.06	&	33.06	&	33.06	&	33.06	&	34.42	&	33.04	&	33.33	\\
	&		&	(0.30\%)	&	(0.001\%)	&	(0.000\%)	&	(0.002\%)	&	(0.006\%)	&	(0.001\%)	&	(4.0\%)	&	(0.07\%)	&	(0.8\%)	\\
1	&	2.73	&	31.31	&	31.31	&	31.31	&	31.31	&	31.31	&	31.31	&	32.26	&	31.31	&	31.48	\\
	&		&	(0.12\%)	&	(0.001\%)	&	(0.001\%)	&	(0.005\%)	&	(0.012\%)	&	(0.002\%)	&	(3.0\%)	&	(0.01\%)	&	(0.5\%)	\\
1.1	&	2.34	&	30.50	&	30.50	&	30.50	&	30.50	&	30.50	&	30.51	&	31.16	&	30.51	&	30.52	\\
	&		&	(0.14\%)	&	(0.000\%)	&	(0.001\%)	&	(0.006\%)	&	(0.009\%)	&	(0.012\%)	&	(2.2\%)	&	(0.02\%)	&	(0.1\%)	\\
\hline
\end{tabular}
\centering
\label{tbl:t0 Sens Z1}
\end{minipage}
\end{table*}

In order to verify that the isotope list we are using is large enough, we calculate with both the NSE6 ($218$ isotopes) and the NSE7 ($260$ isotopes) isotope lists of \citet{Kushnir2019}. The deviations in all cases are negligible. We then test whether the ASE scheme introduces some error by deactivating it (which significantly increases the computational time). Again, the deviations in all cases are negligible. We then test the effects of weak reactions, separately for weak nuclear reactions (with the $\textsc{WEAKLIB}$ module of MESA) and for thermal neutrino emission (with the $\textsc{NEU}$ module of MESA). In order to add weak reactions, we must deactivate the ASE scheme, since it assumes that the plasma reaches equilibrium, which does not hold when weak interactions are included. We find that the effect of weak reactions is negligible in all cases. 

We test the effect of Coulomb corrections to the EOS by repeating the calculations without these corrections. This changes $t_0$ by a few percent and $\nimass$ by $\mysim10\%$ (tens of percent) for high (low) $\nimass$ values. Since Coulomb corrections are known to at least a $10\%$ degree of accuracy \citep[see detailed discussion in][]{Kushnir2019}, the uncertainty because of the Coulomb corrections to the EOS is sub-percent for $t_0$ and a few percent at most for $\nimass$. We test the effect of the nuclear excitation energy contribution to the EOS by repeating the calculations without this contribution. We find this factor to have a negligible effect on $t_0$ and a sub-percent (a few percent) effect on $\nimass$ for high (low) $\nimass$ values, such that this uncertainty is much smaller than the tension of the calculations with the observations. We finally test the effect of the nuclear reaction screening by repeating the calculations without the screening. This has a sub-percent effect on $t_0$, and a few percent ($10-20\%$) effect on $\nimass$ for high (low) $\nimass$ values. Once again, this uncertainty is smaller than the tension of the calculations with the observations. 

We next test the sensitivity of the FLASH simulations to a few numerical choices. We choose for each $\MWD$ and metallicity ($Z=0$ and $Z=Z_{\odot}$) a FLASH maximal resolution for which the deviation of the results ($\nimass$ and $t_0$) from the converged values is $\lesssim5\%$. For each case, we calculate with FLASH the sensitivity of the results to a few numerical choices. The $Z=Z_{\odot}$ results for $\nimass$ and $t_0$ are presented in Table~\ref{tbl:Ni56 Sens Z1 FLASH}  (similar results are presented for $Z=0$ in Appendix~\ref{sec:more results}, Table~\ref{tbl:Ni56 Sens Z0 FLASH}). We test whether the ASE scheme introduces some error by deactivating it (which significantly increases the computational time). The deviations in $\nimass$ ($t_0$) are $\lesssim2\%$ ($\lesssim0.2\%$). We next test the sensitivity for the refinement scheme by increasing the minimal refinement level, $lr_{\min}$ (the default level is $8$, see Section~\ref{sec:FLASH}). For the $\MWD=1,1.1\,M_{\odot}$ cases, we are able to increase $lr_{\min}$ to the maximal refinement level, $lr_{\max}$ (i.e., the calculations did not include mesh refinement at all), and for the other cases, we used $lr_{\min}=12<lr_{\max}$. The $\nimass$ and $t_0$ deviations in all cases were $<10^{-3}$, demonstrating the effectiveness of our refinement scheme. Finally, we calculate for the $\MWD=0.8,0.85,0.9\,M_{\odot}$ cases with $lr_{\min}=lr_{\max}$ up to the time when the shock wave reaches the surface of the WD. We compare the total amount of $^{56}$Ni synthesized up to this time between these runs and our default runs, and we find the deviations to be $\lesssim5\times10^{-3}$. 

\begin{table*}
\begin{minipage}{160mm}
\caption{Sensitivity of the calculated $\nimass$ and $t_0$ of the $Z=Z_{\odot}$ case in FLASH to the ASE scheme. We choose for each $\MWD$ (1st column) a FLASH maximal resolution (2nd column, the corresponding $lr_{\max}$ in the 3rd column) for which the deviation of the results from the converged values is $\lesssim5\%$ (4th and 6th column, the deviation in parenthesis). For each case, we calculate with FLASH without using the ASE scheme (5th and 7th columns, deviation from the reference case in parenthesis).}
\begin{tabular}{|c||c||c||c||c||c||c|}
\hline
$\MWD\,[M_{\odot}]$  & $\Delta x\,[\textrm{km}]$ & $lr_{\max}$  & $\nimass\,[M_{\odot}]$  &  $\nimass\,[M_{\odot}]$   &  $t_0\,[\textrm{day}]$ &  $t_0\,[\textrm{day}]$    \\
 & &  & Reference & w/o ASE & Reference & w/o ASE  \\ \hline
0.8	&	0.125	&	17	&	0.031	&	0.031	&	35.0	&	35.0	\\
     	&		&		&	(1.49\%)	&	(0.005\%)	&	(0.15\%)	&	(0.001\%)	\\
     0.85	&	0.25	&	16	&	0.124	&	0.123	&	34.4	&	34.3	\\
     	&		&		&	(2.16\%)	&	(1.459\%)	&	(0.09\%)	&	(0.032\%)	\\
     0.9	&	0.5	&	15	&	0.256	&	0.252	&	33.0	&	33.0	\\
     	&		&		&	(2.13\%)	&	(1.796\%)	&	(0.22\%)	&	(0.094\%)	\\
     1	&	2	&	13	&	0.531	&	0.525	&	31.4	&	31.4	\\
     	&		&		&	(2.14\%)	&	(1.075\%)	&	(0.44\%)	&	(0.192\%)	\\
     1.1	&	4	&	12	&	0.774	&	0.770	&	30.8	&	30.8	\\
     	&		&		&	(2.59\%)	&	(0.483\%)	&	(1.11\%)	&	(0.132\%)	\\
\hline
\end{tabular}
\centering
\label{tbl:Ni56 Sens Z1 FLASH}
\end{minipage}
\end{table*}

In conclusion, none of the physical and numerical uncertainties tested in this section are significant enough to relieve the tension with the observations. 

\subsection{Sensitivity to the initial heavy element abundances}
\label{sec:initial heavy}

In our calculations, we assume that we can parameterise the heavy element traces of the WD initial composition with $^{22}$Ne alone. Here we show that for our purposes, several other prescriptions are equivalent, if compared at the same $Y_e$. For example, \citet{Shen2018} included $^{56}$Fe in addition to $^{22}$Ne with $X(^{56}\textrm{Fe})=0.1X(^{22}\textrm{Ne})$, and other works included more isotopes, according to solar abundances \citep[e.g.,][]{Blondin2017,Miles2019}. In order to account for more isotopes with solar abundance ratios, while keeping $Y_e$ from Equation~\eqref{eq:Ne22sol}, we define $g$ as:
\begin{eqnarray}\label{eq:g}
Y\left(^{22}\textrm{Ne}\right)&=&g\left[Y_{0}\left(^{12}\textrm{C}\right)+Y_{0}\left(^{14}\textrm{N}\right)+Y_{0}\left(^{16}\textrm{O}\right)\right],\nonumber\\
Y_{i}&=&g Y_{i,0},
\end{eqnarray}
where the $0$ subscript denotes the present-day solar photosphere (with the element abundances of \citet{Asplund2009} and the isotopic fractions of \citet{Lodders2003}), and the index $i$ runs over the additional stable isotopes of all the elements heavier than nitrogen, other than $^{16}$O and $^{22}$Ne. We assume that the isotopes of all elements lighter than oxygen (and $^{16}$O) have converted to $^{12}$C, $^{16}$O and $^{22}$Ne, while the other isotopes are not affected. For this composition, we have
\begin{eqnarray}\label{eq:calc g}
Y_{e}=10Y\left(^{22}\textrm{Ne}\right)+\sum_{i}Z_{i}Y_{i}+\frac{1}{2}\left[1-22Y\left(^{22}\textrm{Ne}\right)+\sum_{i}A_{i}Y_{i}\right],&&\nonumber\\
\Rightarrow g=\frac{1-2 Y_{e}}{2\left[Y_{0}\left(^{12}\textrm{C}\right)+Y_{0}\left(^{14}\textrm{N}\right)+Y_{0}\left(^{16}\textrm{O}\right)\right]+\sum_{i}Y_{i,0}\left(A_{i}-2Z_{i}\right)},&&
\end{eqnarray}
which completely defines the initial composition. 

For example, assume we want to add $^{56}$Fe, with $Y_0(^{56}$Fe$)\approx2.12\times10^{-5}$. For solar metallicity, we find that $g\approx1.054$, $X(^{12}\textrm{C})+X(^{16}\textrm{O})\approx0.9847$, $X(^{22}\textrm{Ne})\approx0.0140$ and $X(^{56}\textrm{Fe})\approx0.0013$. As another example, we add all stable isotopes (of all the elements heavier than nitrogen, other than $^{16}$O and $^{22}$Ne) that are included in our default $178$-isotope list. We find that $\sum_{i}A_{i}Y_{i,0}\approx5.09\times10^{-3}$ and $\sum_{i}Z_{i}Y_{i,0}\approx2.48\times10^{-3}$ (in cases where only part of the stable isotopes of some element are included in the isotope list, we renormalise their fraction such that they will sum to unity), with $X(^{12}\textrm{C})+X(^{16}\textrm{O})\approx0.9817$,  $X(^{22}\textrm{Ne})\approx0.0137$ and $\sum_i X_i\approx0.0046$. 

We use the same resolutions from Section~\ref{sec:uncertainty} to calculate for each $\MWD$ and $Z=Z_{\odot}$ with four different initial compositions, while keeping the same $Y_e$. The deviations of $\nimass$ and $t_0$ are shown in Tables~\ref{tbl:Ni56 Comp Sens} and~\ref{tbl:t0 Comp Sens}, respectively. The different initial compositions include $X(^{22}\rm{Ne})=0.015$, $X(^{12}\rm{C})=0.485(0.5)$, $X(^{16}\rm{O})=0.5(0.485)$,
and the two examples from above. We find deviations of up to a few percent in $\nimass$ and less than $0.5\%$ in $t_0$. We conclude that this uncertainty is smaller than the tension of the calculations with the observations. 

\begin{table*}
\begin{minipage}{160mm}
\caption{Sensitivity of the calculated $\nimass$ (in $M_{\odot}$) of the $Z=Z_{\odot}$ case to the initial heavy element abundances. We use the same reference calculations as that in Table~\ref{tbl:Ni56 Sens Z1} (1st-3rd columns). For each case, we calculate with V1D four different initial compositions, while keeping the same $Y_e$, in the 4th-7th columns (deviation from the reference case in parenthesis). 4th column: $X(^{12}\rm{C})=0.485$, $X(^{16}\rm{O})=0.5$, $X(^{22}\rm{Ne})=0.015$. 5th column: $X(^{12}\rm{C})=0.5$, $X(^{16}\rm{O})=0.485$, $X(^{22}\rm{Ne})=0.015$. 6th column: $X(^{12}\textrm{C})=X(^{16}\textrm{O})=0.4924$, $X(^{22}\textrm{Ne})=0.0140$ and $X(^{56}\textrm{Fe})=0.0012$. 7th column: $X(^{12}\textrm{C})=X(^{16}\textrm{O})=0.4908$, $X(^{22}\textrm{Ne})=0.0137$ and $\sum_i X_i=0.0047$ (see text).}
\begin{tabular}{|c||c||c||c||c||c||c|}
\hline
$\MWD\,[M_{\odot}]$  & $\Delta x_0\,[\textrm{km}]$ & Reference & $X(^{16}\rm{O})=0.5$  & $X(^{12}\rm{C})=0.5$  & $X(^{56}\textrm{Fe})=0.1X(^{22}\textrm{Ne})$ & all stable isotopes \\ \hline
0.8	&	0.87	&	0.026	&	0.025	&	0.027	&	0.026	&	0.026	\\
	&		&	(3.7\%)	&	(2.4\%)	&	(2.5\%)	&	(1.3\%)	&	(0.3\%)	\\
0.85	&	0.82	&	0.114	&	0.109	&	0.119	&	0.113	&	0.110	\\
	&		&	(6.0\%)	&	(4.3\%)	&	(4.7\%)	&	(0.8\%)	&	(3.4\%)	\\
0.9	&	1.55	&	0.251	&	0.249	&	0.253	&	0.250	&	0.249	\\
	&		&	(3.1\%)	&	(0.8\%)	&	(0.8\%)	&	(0.1\%)	&	(0.7\%)	\\
1	&	2.73	&	0.533	&	0.532	&	0.535	&	0.533	&	0.532	\\
	&		&	(1.0\%)	&	(0.3\%)	&	(0.3\%)	&	(0.0\%)	&	(0.3\%)	\\
1.1	&	2.34	&	0.791	&	0.790	&	0.792	&	0.791	&	0.791	\\
	&		&	(0.2\%)	&	(0.1\%)	&	(0.1\%)	&	(0.0\%)	&	(0.1\%)	\\
\hline
\end{tabular}
\centering
\label{tbl:Ni56 Comp Sens}
\end{minipage}
\end{table*}

\begin{table*}
\begin{minipage}{160mm}
\caption{Same as Table~\ref{tbl:Ni56 Comp Sens} for $t_0$ (in day).}
\begin{tabular}{|c||c||c||c||c||c||c|}
\hline
$\MWD\,[M_{\odot}]$  & $\Delta x_0\,[\textrm{km}]$ & Reference & $X(^{16}\rm{O})=0.5$  & $X(^{12}\rm{C})=0.5$  & $X(^{56}\textrm{Fe})=0.1X(^{22}\textrm{Ne})$ & all stable isotopes \\ \hline
0.8	&	0.87	&	34.9	&	35.0	&	34.8	&	34.9	&	34.9	\\
	&		&	(0.2\%)	&	(0.3\%)	&	(0.3\%)	&	(0.1\%)	&	(0.1\%)	\\
0.85	&	0.82	&	34.5	&	34.6	&	34.3	&	34.5	&	34.6	\\
	&		&	(0.3\%)	&	(0.5\%)	&	(0.5\%)	&	(0.1\%)	&	(0.4\%)	\\
0.9	&	1.55	&	33.1	&	33.2	&	33.0	&	33.1	&	33.1	\\
	&		&	(0.3\%)	&	(0.3\%)	&	(0.3\%)	&	(0.1\%)	&	(0.3\%)	\\
1	&	2.73	&	31.3	&	31.4	&	31.2	&	31.3	&	31.4	\\
	&		&	(0.1\%)	&	(0.2\%)	&	(0.2\%)	&	(0.1\%)	&	(0.2\%)	\\
1.1	&	2.34	&	30.5	&	30.6	&	30.4	&	30.5	&	30.6	\\
	&		&	(0.1\%)	&	(0.2\%)	&	(0.2\%)	&	(0.1\%)	&	(0.2\%)	\\
\hline
\end{tabular}
\centering
\label{tbl:t0 Comp Sens}
\end{minipage}
\end{table*}

\subsection{Sensitivity to the initial WD profile}
\label{sec:initial profile}

The initial profiles of the WDs that we have considered so far include a few simplifying assumptions: The WDs were isothermal (with $T_{\textrm{WD},9}=0.01$), the initial composition was uniform, and the mass fractions of $^{12}$C and $^{16}$O were roughly equal. However, evolutionary models of WDs suggest that modifications to these assumptions are required \citep[see e.g.,][]{Renedo2010,Lauffer2018}. In this section, we study the sensitivity of our results to a few of these assumptions. In Section~\ref{sec:C/O ratio}, we keep the assumption of uniform initial composition but allow the mass fraction ratio of $^{12}$C/$^{16}$O (hereafter C/O) to vary. In Section~\ref{sec:TWD}, we keep the isothermal assumption but test the sensitivity to the value of $T_{\textrm{WD}}$. 

\subsubsection{Sensitivity to C/O}
\label{sec:C/O ratio}

\begin{figure}
\includegraphics[width=0.48\textwidth]{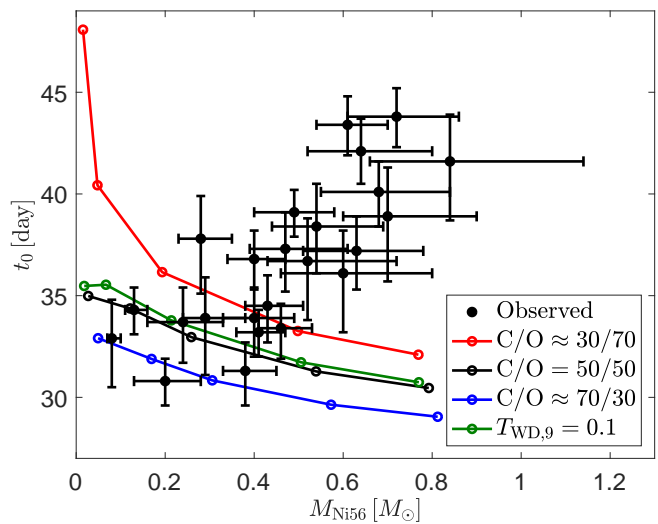}
\caption{The effect of the initial WD profile on the $t_0-\nimass$ relation. The observation (filled circles) and the V1D $Z=Z_{\odot}$ results (black) are the same as in Figure~\ref{fig:Nit0}. Red (blue) line: the converged V1D results (see text for details) for C/O$\myapprox30/70(70/30)$. Green line: the converged V1D results for $T_{\textrm{WD},9}=0.01$. $t_0$ increases for C/O$\myapprox30/70$ for all $\nimass$ values. While the increase for high $\nimass$ values is insufficient to explain the observations, the agreement with the observations for low $\nimass$ values diminishes. For C/O$\myapprox70/30$, $t_0$ decreases for all $\nimass$ values, which increases the tension with the observations. The effect of changing the $T_{\textrm{WD}}$ on the $t_0-\nimass$ relation is small. 
\label{fig:Nit0_C3070}}
\end{figure}

Evolutionary models of WDs suggest that the composition within the star is roughly uniform and bounded between C/O$\myapprox50/50$ and C/O$\myapprox30/70$ \citep{Renedo2010,Lauffer2018}. We therefore test the sensitivity of our V1D $Z=Z_{\odot}$ results to the value of C/O. We calculate two cases: $X(^{12}\rm{C})=0.2925$, $X(^{16}\rm{O})=0.6925$, $X(^{22}\rm{Ne})=0.015$ (C/O$\myapprox30/70$, which corresponds to the smallest $^{12}$C fraction suggested by evolutionary models) and $X(^{12}\rm{C})=0.6925$, $X(^{16}\rm{O})=0.2925$, $X(^{22}\rm{Ne})=0.015$ (C/O$\myapprox70/30$). For each composition and $\MWD$, we perform a convergence test similar to the one in Section~\ref{sec:convergence}, restricting to $f=0.1(0.05)$ for $\MWD=0.8,0.85,0.9(1,1.1)\,M_{\odot}$. The converged results, with the same resolutions of Table~\ref{tbl:Ni56 Conv Z1}, are presented in Figure~\ref{fig:Nit0_C3070}. As can be seen in the figure, $t_0$ increases for C/O$\myapprox30/70$ for all $\nimass$ values. While the increase for high $\nimass$ values is insufficient to explain the observations, the agreement with the observations for low $\nimass$ values diminishes. For C/O$\myapprox70/30$, $t_0$ decreases for all $\nimass$ values, which increases the tension with the observations. 

\subsubsection{Sensitivity to $T_{\textrm{WD}}$}
\label{sec:TWD}

The core temperature of very young WDs is $T_{\textrm{WD},9}\sim0.1$, and they cool down to $T_{\textrm{WD},9}\sim10^{-3}$ by the time they are very old WDs. Our EOS is only valid for $T_{\textrm{WD},9}\gtrsim\rm{few}\times10^{-3}$ (for an initial density of $\mysim10^{7}\,\textrm{g}\,\textrm{cm}^{-3}$), since the ion coupling parameter of the plasma, $\Gamma$, is larger than $200$ for lower temperatures, where the fit for $f(\Gamma)$ is not valid (see Appendix~\ref{sec:input} for details). Although we do not expect significant changes for lower temperatures, we are currently not able to test this, which is also the reason that our default value is $T_{\textrm{WD},9}=0.01$. We are able to test higher temperatures, and here we examine the $T_{\textrm{WD},9}=0.1$ case, relevant for very young WDs. For each WD mass, we perform a convergence test, similar to the one in Section~\ref{sec:convergence}, restricting to $f=0.1(0.05)$ for $\MWD=0.8,0.85,0.9(1,1.1)\,M_{\odot}$. The converged results, with the same resolutions as those in Table~\ref{tbl:Ni56 Conv Z1}, are presented in Figure~\ref{fig:Nit0_C3070}. As can be seen in the figure, the effect of changing the $T_{\textrm{WD}}$ on the $t_0-\nimass$ relation is small. 


\section{A calibration of a 69-isotope network}
\label{sec:small network}

We have shown in Section~\ref{sec:uncertainty} that increasing the number of isotopes has a negligible effect on the calculated $\nimass$ and $t_0$. This result suggests that our 178-isotope list can be significantly reduced while maintaining high accuracy for the calculation of $\nimass$ and $t_0$. A reduced isotopes list decreases the required computational resources and it is essential for multi-D calculations. In this section, we calibrate a 69-isotope list that allows a $\lesssim1\%$ accuracy for the calculation of $\nimass$ and $t_0$ for the $Z=Z_{\odot}$ case. The reduced network includes only $231$ reactions (and their inverse reactions), which allows us to perform in Section~\ref{sec:reaction rate sensitivity} a sensitivity check of our results to the uncertainty of the reaction rate values. 

In order to find the reduced network, we use the following method. We choose for each $\MWD$ a V1D resolution that allows a relatively fast calculation with reasonable accuracy (same as the chosen resolutions is Section~\ref{sec:uncertainty}, except for $\MWD=0.8\,M_{\odot}$ and $\MWD=0.85\,M_{\odot}$, where a factor-two coarser resolution is chosen). We begin with the $178$ isotope list and remove (by an educated guess) one isotope from the list. We calculate with the new list for each WD mass and inspect the deviations in $\nimass$ and $t_0$. For small deviations in all WD masses, we continue with the new list. Otherwise, we return the inspected isotope to the list. We then repeat the process with a different chosen isotope. The process finishes after we have inspected all isotopes. The final list includes $69$ isotopes: $n$, $p$, $^{4}$He, $^{11}$B, $^{12-14}$C, $^{13-15}$N, $^{16-17}$O, $^{17}$F, $^{20-22}$Ne, $^{21,23}$Na, $^{23-26}$Mg, $^{26-27}$Al, $^{27-30}$Si, $^{29,31}$P, $^{32-34}$S, $^{35}$Cl, $^{36,38}$Ar, $^{39}$K, $^{40-44}$Ca, $^{43-45}$Sc, $^{44-47}$Ti, $^{47-49}$V, $^{48-50}$Cr, $^{51-53}$Mn, $^{52-56}$Fe, $^{55-57}$Co, $^{56-58}$Ni. With this list, the deviations in $\nimass$ and $t_0$ are no more than a percent. 

\begin{figure}
\includegraphics[width=0.48\textwidth]{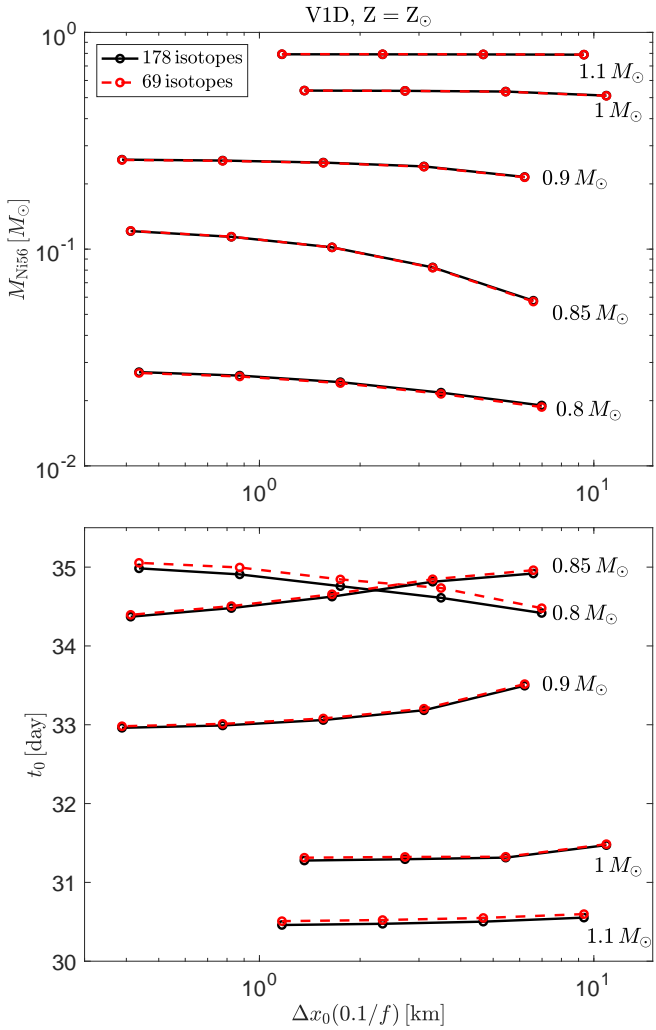}
\caption{Same as Figure~\ref{fig:Ni-t0_Convergence_Z1}, but with only the V1D results with $f=0.1(0.05)$ for $M_{WD}=0.8,0.85,0.9(1,1.1)$ presented. Calculations with the $178$-isotope list (solid black line) are compared to calculations with the $69$-isotope list (dashed red line). The convergence properties of both isotope lists are similar. The deviations of the converged $\nimass$ ($t_0$) values between the two isotope lists are smaller than one percent ($0.2\%$). 
\label{fig:Ni-t0_Convergence_Z1_69}}
\end{figure}

We next present in Figure~\ref{fig:Ni-t0_Convergence_Z1_69} a V1D resolution convergence test (similar to the one performed in Section~\ref{sec:default}) with the $69$-isotope list. As can be seen in the figure, the convergence properties of the $69$-isotope list are very similar to those of the $178$-isotope list. The deviations of the converged $\nimass$ ($t_0$) values between the two isotope lists are smaller than one percent ($0.2\%$). 

\section{Sensitivity of the results to reaction rate uncertainty}
\label{sec:reaction rate sensitivity}

The $69$-isotope network that was calibrated in Section~\ref{sec:small network} includes only $231$ reactions (and their inverse reactions), which allows us to perform in this section a sensitivity check of our results to the uncertainty of the reaction rate values. We conduct $231$ sets of calculations, where for each set we perform V1D calculations for all $\MWD$ values (with the resolutions of Section~\ref{sec:small network}) and multiply one reaction (and its inverse) by a (temperature-independent) factor $f_r=0.5,2$ ($10$ calculations in each set). This procedure allows us to find the reactions that control the uncertainty of our results. For most reactions, the changes of $\nimass$ and $t_0$ are smaller than one percent, making them unlikely to affect the uncertainty budget. We then group the rest of the reactions according to the changes in $\nimass$ and $t_0$. The $29$ reactions that only change the $\nimass$ of $\MWD=0.8\,M_{\odot}$ and/or $\MWD=0.85\,M_{\odot}$ by more than a percent (and by less than $6$ percent) are in group I. The other $13$ reactions change also the $\nimass$ of the other WD masses and/or $t_0$ by more than a percent, and are in group II. The reactions in group I and II are listed in Table~\ref{tbl:reactions}. It is evident that the reactions in Group II mostly involve elements with $A\lesssim24$. These reactions are related to the inverse triple-$\alpha$ bottleneck that controls the approach to NSE, which determines the length scale of the TNDW \citep{Khokhlov1989,Kushnir2019}.

\begin{table*}
\caption{Reactions that control the error budget of our results, calculated with the $69$-isotope network ($231$ reactions and their inverses). Group I (1st column) includes $29$ reactions that only change the $\nimass$ of $\MWD=0.8\,M_{\odot}$ and/or $\MWD=0.85\,M_{\odot}$ by more than a percent (and by less than $6$ percent). Group II (2nd column) includes $13$ reactions that change also the $\nimass$ of the other WD masses and/or $t_0$ by more than a percent. Highlighted reactions have no uncertainty estimate in version \textsc{v65a\_090817} of STARLIB \citep{Sallaska2013}.}
\begin{tabular}{|c||c|}
\hline
Group I  &    Group II \\ \hline
$^{12}$C$(n,\gamma)^{13}$C & $^{23}$Na$(p,\gamma)^{24}$Mg	\\
$^{20}$Ne$(n,\gamma)^{21}$Ne & \bf{$\bf{^{21}}$Na$\boldsymbol{(\alpha,p)^{24}}$Mg}	\\
$^{32}$S$(n,\gamma)^{33}$S & \bf{$\bf{^{13}}$N$\boldsymbol{(\alpha,p)^{16}}$O}	\\
$^{29}$Si$(n,\gamma)^{30}$Si & \bf{$\bf{^{23}}$Na$\boldsymbol{(\alpha,p)^{26}}$Mg} 	\\
\bf{$\bf{^{44}}$Ti$\boldsymbol{(n,\gamma)^{45}}$Ti} &  \bf{$\bf{^{23}}$Na$\boldsymbol{(\alpha,n)^{26}}$Al}	\\
\bf{$\bf{^{23}}$Na$\boldsymbol{(p,n)^{23}}$Mg} & $^{20}$Ne$(\alpha,\gamma)^{24}$Mg	\\
$^{21}$C$(p,n)^{21}$Na & $^{16}$O$(\alpha,\gamma)^{20}$Ne	\\
$^{39}$K$(p,\gamma)^{40}$Ca & $^{12}$C$(\alpha,\gamma)^{16}$O	\\
$^{26}$Mg$(p,\gamma)^{27}$Al & $^{12}$C$+\gamma \leftrightarrow 3\alpha$		\\
$^{20}$Ne$(p,\gamma)^{21}$Na & 	\bf{$\bf{^{12}}$C$\boldsymbol{+^{12}}$C$\boldsymbol{\leftrightarrow p+^{23}}$Na} \\
\bf{$\bf{^{44}}$Sc$\boldsymbol{(p,\gamma)^{45}}$Ti} & 	\bf{$\bf{^{12}}$C$\boldsymbol{+^{12}}$C$\boldsymbol{\leftrightarrow \alpha+^{20}}$Ne} \\
\bf{$\bf{^{45}}$Sc$\boldsymbol{(p,\gamma)^{46}}$Ti} & \bf{$\bf{^{12}}$C$\boldsymbol{+^{16}}$O$\boldsymbol{\leftrightarrow \alpha+^{24}}$Mg}	\\
$^{30}$Si$(n,\gamma)^{31}$P & 	\bf{$\bf{^{12}}$C$\boldsymbol{+^{16}}$O$\boldsymbol{\leftrightarrow p+^{27}}$Al} \\
\bf{$\bf{^{27}}$Al$\boldsymbol{(\alpha,p)^{30}}$Si} & 	\\
\bf{$\bf{^{42}}$Ca$\boldsymbol{(\alpha,p)^{45}}$Sc} & 	\\
\bf{$\bf{^{17}}$F$\boldsymbol{(\alpha,p)^{20}}$Ne} & 	\\
\bf{$\bf{^{23}}$Mg$\boldsymbol{(\alpha,p)^{26}}$Al} & 	\\
$^{20}$Ne$(\alpha,p)^{23}$Na & 	\\
\bf{$\bf{^{44}}$Ti$\boldsymbol{(\alpha,p)^{47}}$V} & 	\\
$^{13}$C$(\alpha,n)^{16}$O & 	\\
$^{26}$Mg$(\alpha,n)^{29}$Si & 	\\
\bf{$\bf{^{42}}$Ca$\boldsymbol{(\alpha,n)^{45}}$Ti} & 	\\
\bf{$\bf{^{20}}$Ne$\boldsymbol{(\alpha,n)^{23}}$Mg} & 	\\
$^{17}$O$(\alpha,n)^{20}$Ne & 	\\
\bf{$\bf{^{11}}$B$\boldsymbol{(\alpha,n)^{14}}$N} & 	\\
\bf{$\bf{^{42}}$Ca$\boldsymbol{(\alpha,\gamma)^{46}}$Ti} & 	\\
\bf{$\bf{^{12}}$C$\boldsymbol{+^{12}}$C$\boldsymbol{\leftrightarrow n+^{23}}$Mg}	 & \\
\bf{$\bf{^{16}}$O$\boldsymbol{+^{16}}$O$\boldsymbol{\leftrightarrow p+^{31}}$P}	 & 	\\
\bf{$\bf{^{12}}$C$\boldsymbol{+^{16}}$O$\boldsymbol{\leftrightarrow n+^{27}}$Si}	 & 	\\
\hline
\end{tabular}
\centering
\label{tbl:reactions}
\end{table*}

\begin{figure}
\includegraphics[width=0.48\textwidth]{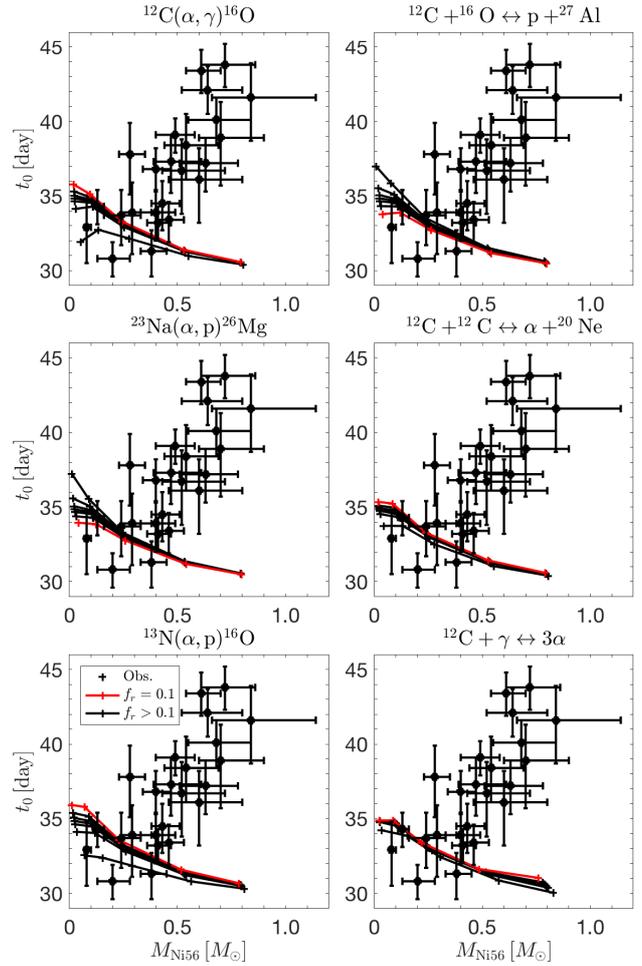}
\caption{A few examples of the effect of the reaction rate uncertainty on the $t_0-\nimass$ relation. Each panel is the same as Figure~\ref{fig:Nit0}, but with only the $Z=Z_{\odot}$ V1D results plotted (see text for the resolutions of the calculations). In each panel, one reaction (and its inverse) is multiplied by a (temperature-independent) factor $f_r=0.1,0.5,0.8,1,1.25,2,10$. Calculations with the same $f_r$ value are connected with a black line (red line for $f_r=0.1$). The tension between the predictions of this model and the observed $t_0-\nimass$ relation for high-luminosity ($\nimass\gtrsim0.5\,M_{\odot}$) SNe Ia is much larger than the uncertainty of the results.  
\label{fig:Nit0_RatesSens}}
\end{figure}

We highlight in Table~\ref{tbl:reactions} the reactions for which no uncertainty estimate is provided by version \textsc{v65a\_090817} of STARLIB\footnote{https://starlib.github.io/Rate-Library/} \citep{Sallaska2013}. The reactions that belong to group I can contribute $20-30\%$ to the $\nimass$ uncertainty of $\MWD=0.8\,M_{\odot}$ and/or $\MWD=0.85\,M_{\odot}$ (for $f_r=0.5,2$). This is comparable to the uncertainty of some single group II reactions (that do not have an uncertainty estimate), so we focus in what follows on group II reactions. We perform more sets of calculations for group II reactions with $f_r=0.1,0.8,1.25,10$. The combined uncertainty from the reactions in this group (for $f_r=0.1,10$), can change $\nimass$ by a factor of few ($\mysim10\%$) and $t_0$ by $\mysim10\%$ ($1-2\%$) for $\MWD=0.8,0.85\,M_{\odot}$ ($\MWD=0.9,1,1.1\,M_{\odot}$). The available uncertainty estimate for $5$ reactions in this group can somewhat decrease the combined uncertainty.  We provide in Figure~\ref{fig:Nit0_RatesSens} a few examples for the effect of the reaction rate uncertainty on the $t_0-\nimass$ relation. As can be seen in the figure, the tension between the predictions of this model and the observed $t_0-\nimass$ relation is much larger than the uncertainty of the results.  
 
\section{Comparison to previous works}
\label{sec:comparison}

\begin{figure}
\includegraphics[width=0.48\textwidth]{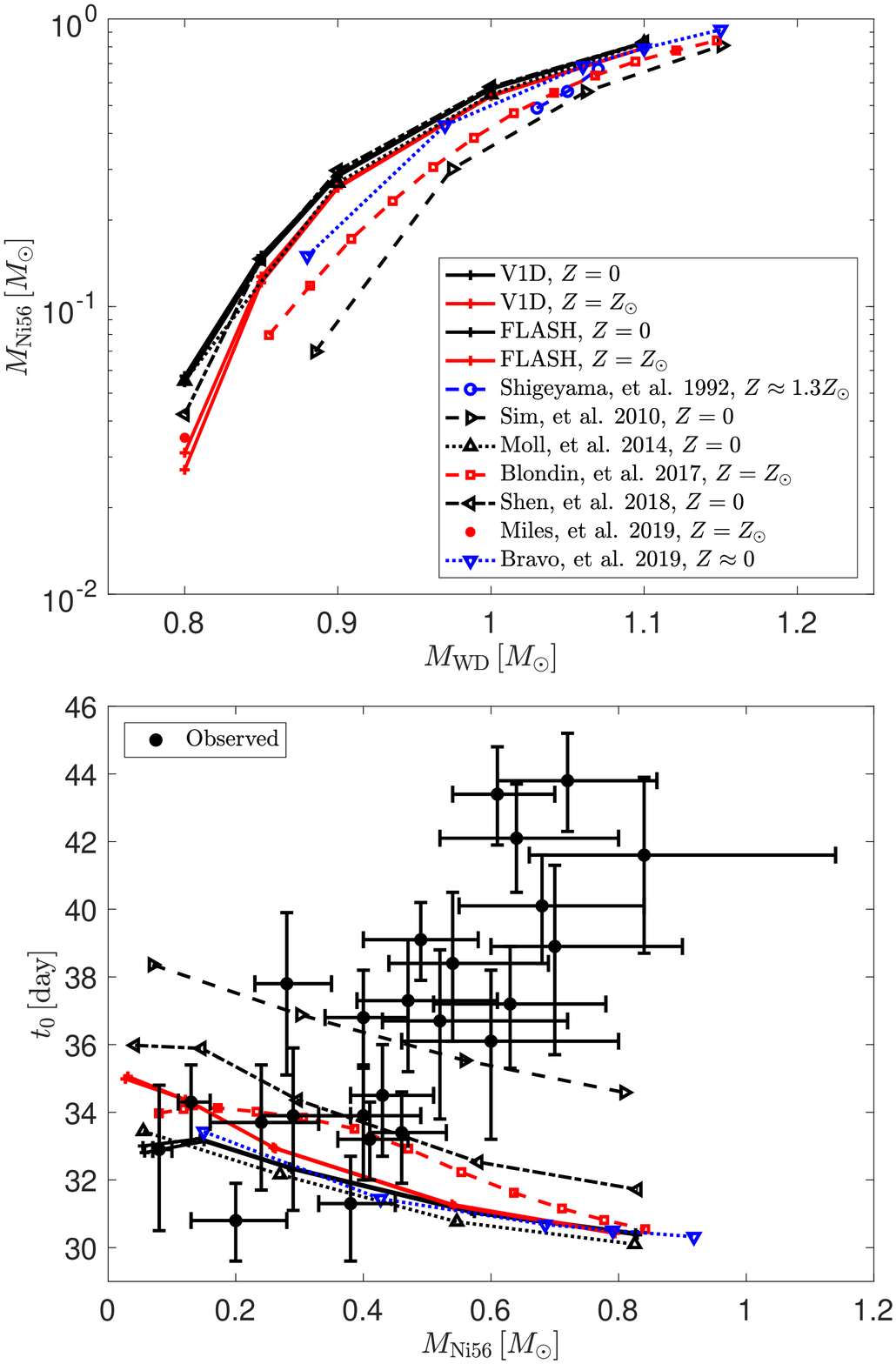}
\caption{The calculated $\nimass$ values as a function of the $\MWD$ (top panel) and the $t_{0}-\nimass$ relation (bottom panel) in different studies of SCD. Plus signs: the converged $Z=0$ (black) and $Z=Z_{\odot}$ (red) results calculated with V1D and FLASH (this work). Circles connected with dashed blue line: the $Z\approx1.3Z_{\odot}$ results of \citet[][only top panel]{Shigeyama1992}. Right-pointing triangles connected with dashed black lines: the $Z=0$ results of \citet{Sim2010}. Upward-pointing triangles connected with dotted black lines: the $Z=0$ results of \citet{Moll2014}. Squares connected with dashed red lines: the $Z=Z_{\odot}$ results of \citet{Blondin2017}. Left-pointing triangles connected with dashed black lines: the $Z=0$ results of \citet{Shen2018}. Red filled circle: the $Z=Z_{\odot}$ result of \citet[][only top panel]{Miles2019}. Downward-pointing triangles connected with dotted blue lines: the $Z\approx0$ results of \citet{Bravo2019}. The observed sample is the same as in Figure~\ref{fig:Nit0}. The general $\nimass-\MWD$ and $t_0-\nimass$ relations are reproduced in all previous works \citep[except for the results of][which are systematically different from all other works, see Section~\ref{sec:Sim 2010}]{Sim2010}. Specifically, the tension with the observed $t_0-\nimass$ relation exists in all previous studies.
\label{fig:ComparePrevious}}
\end{figure}

In this section, we compare our results to previous studies of SCD, performed with less accurate numerical schemes. Each subsection contains a careful comparison to one previous work \citep[we thank the authors of][for sharing their ejecta profiles with us]{Sim2010,Moll2014,Blondin2017,Shen2018,Bravo2019}, with a focus on comparing $\nimass$, which is more sensitive and easier to compare than $t_0$. We find that the general $\nimass-\MWD$ and $t_0-\nimass$ relations (Figure~\ref{fig:ComparePrevious}) are reproduced in all previous works \citep[except for the results of][which are systematically different from all other works, see Section~\ref{sec:Sim 2010}]{Sim2010}. Specifically, the tension with the observed $t_0-\nimass$ relation exists in all previous studies. The differences between previous works and our results are discussed in detail.

\subsection{Comparison to \citet{Shigeyama1992}}
\label{sec:Shigeyama 1992}

\citet{Shigeyama1992} used a Lagrangian PPM code \citep{Colella1984,Colella1985} to calculate SCD. They provide $\nimass$ values for WD masses in the range of $1.03-1.07\,M_{\odot}$ with a uniform composition of $X(^{12}$C$)=0.48$, $X(^{16}$O$)=0.5$, and $X(^{22}$Ne$)=0.02$, which corresponds to $Z\approx1.3Z_{\odot}$ with our definition of solar metallicity (see Section~\ref{sec:initial}). The hydrodynamical calculations contained a $13$-isotope $\alpha$-network and post-processing with a $299$-isotope network. The $\nimass$ values obtained by \citet{Shigeyama1992} are compared in Figure~\ref{fig:ComparePrevious} (circles connected with dashed blue line) to our default cases (solid red line). There is reasonable agreement between the results. Because of the limited information we have regarding the calculations of \citet{Shigeyama1992}, we do not attempt here a more detailed comparison. 

\subsection{Comparison to \citet{Sim2010}}
\label{sec:Sim 2010}

\citet{Sim2010} used PROMETHEUS \citep[Eulerian code, see details in][]{Fink2007} to calculate SCD. WD masses in the range of $0.81-1.15\,M_{\odot}$, mostly with zero metallicity, were considered. The hydrodynamical calculations contained cell sizes of $\Delta x\approx10-17\,\textrm{km}$ and a $4$-isotope network, while the location of the TNDW is pre-determined with the level-set technique \citep{Reinecke1999}. This technique assumes that the TNDW propagates with the steady-state solution, regardless of the actual conditions before and after the wave and regardless of the numerical resolution. This situation makes the meaning of convergence tests for this technique somewhat vague. Tracer particles were included for post-processing with a $383$-isotope network. They used $T_{\rm{WD},9}=5\times10^{-4}$, which is too low to be correctly described by the EOS  of \citet{Sim2010}, see details in Section~\ref{sec:TWD}.    

We calculate the cases $\MWD=0.88,1.06\,M_{\odot}$ with zero metallicity, studied by \citet{Sim2010}, using the same initial setup. We use both the input physics of \citet{Sim2010} and our default input physics. In order to match the input physics of \citet{Sim2010}, we did not use Coulomb corrections and we did not include the nuclear excitation correction to the EOS. We calculated with our default $178$-isotope list and $T_{\rm{WD},9}=0.03$. 

\begin{figure}
\includegraphics[width=0.48\textwidth]{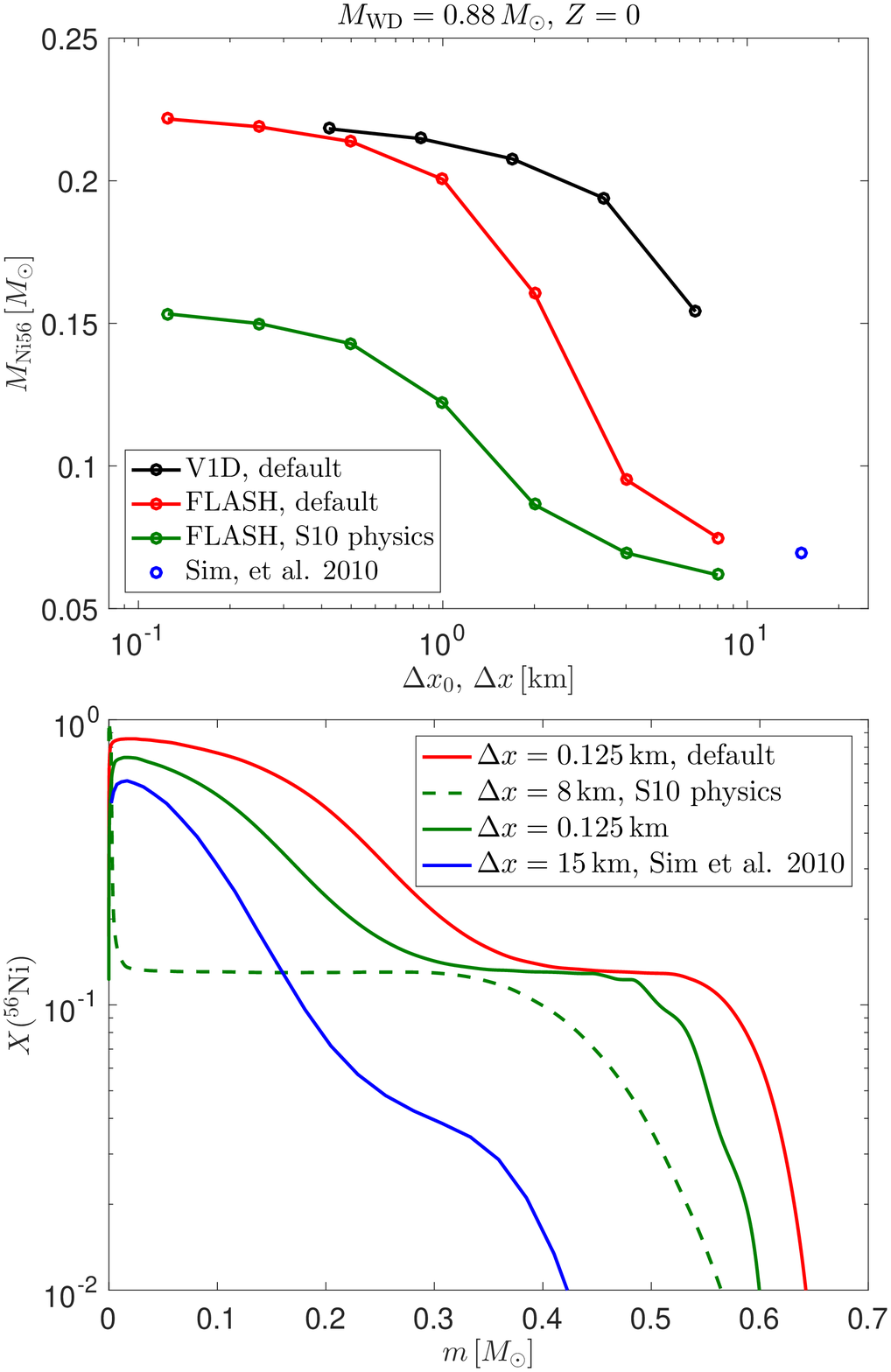}
\caption{Comparison to the $\MWD=0.88\,M_{\odot}$, $Z=0$ case of \citet{Sim2010}. Top panel: $\nimass$ as a function of the V1D initial (FLASH maximal) resolution, $\Delta x_0$ ($\Delta x$). Black (red) line: V1D (FLASH) results with our default input physics. Green line: FLASH results with \citet{Sim2010} input physics, Blue circle: the result of \citet{Sim2010}. Bottom panel: the $^{56}$Ni mass fraction distribution, $X(^{56}\rm{Ni})$, within the  ejecta, as a function of $m$. Red line: the converged ($\Delta x=0.125\,\textrm{km}$) FLASH result with our default input physics. Green lines: The low resolution ($\Delta x=8\,\textrm{km}$, dashed line) and converged ($\Delta x=0.125\,\textrm{km}$, solid line) FLASH results with \citet{Sim2010} input physics. Blue line: the result of \citet{Sim2010}. Our results in the top panel are similar to the results of \citet{Sim2010} when compared at the same resolution. However, this similarity is accidental, since the $^{56}$Ni mass profiles within the ejecta are very different (compare the dashed-green and blue lines in the bottom panel). While the results of our scheme at such low resolution are far from the converged results, the scheme of \citet{Sim2010} is forcing the TNDW to propagate at some predetermined velocity, which leads to a reasonable profile (compare to the converged FLASH profile, solid green line in the bottom panel). Nevertheless, the converged FLASH $\nimass$ value is higher by $\myapprox75\%$  than the value of \citet{Sim2010}. Using our default physics input increases the converged FLASH $\nimass$ by $\myapprox35\%$, mostly because of the inclusion of Coulomb corrections. The V1D $\nimass$ converged value is $\myapprox1.6\%$ higher than the FLASH converged value. 
\label{fig:CompareSimEjecta088}}
\end{figure}

The $\nimass$ values that we obtained for $\MWD=0.88\,M_{\odot}$ with the input physics of \citet{Sim2010} (green line, FLASH with $f=0.1$) are compared to the results of \citet{Sim2010} (blue circle) in the top panel of Figure~\ref{fig:CompareSimEjecta088}. As can be seen in the figure, our results are similar to the results of \citet{Sim2010} when compared at the same resolution. However, this similarity is accidental, since the $^{56}$Ni mass profiles within the ejecta are very different; see the bottom panel of Figure~\ref{fig:CompareSimEjecta088} (compare the dashed green and blue lines). While the results of our scheme at such low resolution are far from the converged results, the scheme of \citet{Sim2010} is forcing the TNDW to propagate at some predetermined velocity, which leads to a reasonable profile (compare to the converged FLASH profile, solid green line in the bottom panel). Nevertheless, the converged FLASH $\nimass$ value is higher by $\myapprox75\%$  than the value of \citet{Sim2010}, and the scheme of \citet{Sim2010} does not allow a proper convergence study. The $^{56}$Ni mass profile of \citet{Sim2010} is more concentrated than the converged FLASH profile, which leads to a $t_0$ value that is higher by a few days. Using our default physics input increases the converged FLASH $\nimass$ by $\myapprox35\%$ (red solid lines in Figure~\ref{fig:CompareSimEjecta088}), mostly because of the inclusion of Coulomb corrections (that also change the initial WD profiles, see Table~\ref{tbl:Ni56 Sens Z0}). This also has the effect of decreasing $t_0$ by roughly a day, see Table~\ref{tbl:t0 Sens Z0}. We further calculate with V1D using our default input physics (black line in the top panel of Figure~\ref{fig:CompareSimEjecta088}, we use $f=0.1$) and we find that the $\nimass$ converged value is higher by $\myapprox1.6\%$ than the FLASH converged value, which is similar to the comparison of Section~\ref{sec:convergence}. 

\begin{figure}
\includegraphics[width=0.48\textwidth]{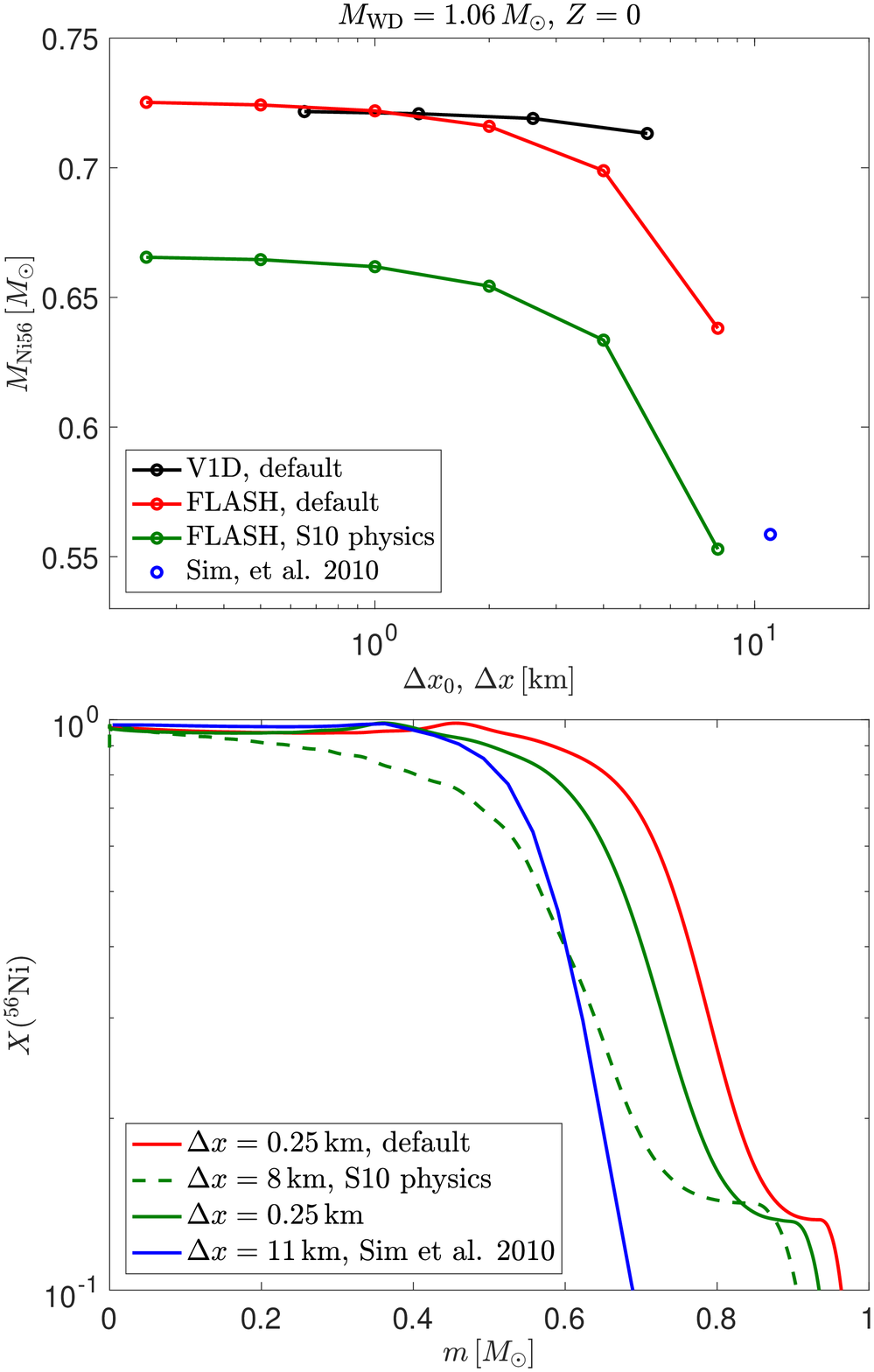}
\caption{Same as Figure~\ref{fig:CompareSimEjecta088} for $\MWD=1.06\,M_{\odot}$. Our results are similar to the results of \citet{Sim2010} when compared at the same resolution. Again, this similarity is accidental, since the $^{56}$Ni mass profiles within the ejecta are quite different (compare the dashed green and blue lines in the bottom panel). The converged FLASH $\nimass$ value is $\myapprox15\%$ higher than the value of \citet{Sim2010}. Using our default input physics increases the converged FLASH $\nimass$ by $\myapprox10\%$ (red lines), mostly because of the inclusion of Coulomb corrections (see Table~\ref{tbl:Ni56 Sens Z0}). The V1D $\nimass$ converged value deviates by $\myapprox0.5\%$ from the FLASH converged value.
\label{fig:CompareSimEjecta106}}
\end{figure}

The values of $\nimass$ that we obtained for $\MWD=1.06\,M_{\odot}$ with the input physics of \citet{Sim2010} (green line, FLASH with $f=0.1$) are compared to the results of \citet{Sim2010} (blue circle) in the top panel of Figure~\ref{fig:CompareSimEjecta106}. As can be seen in the figure, our results are similar to the results of \citet{Sim2010} when compared at the same resolution. Again, this similarity is accidental, since the $^{56}$Ni mass profiles within the ejecta are quite different, see the bottom panel of Figure~\ref{fig:CompareSimEjecta106} (compare the dashed green and blue lines). The converged FLASH $\nimass$ value is $\myapprox15\%$ higher than the value of \citet[][see the converged profile in solid green line in the bottom panel]{Sim2010}. Again, the $^{56}$Ni mass profile of \citet{Sim2010} is more concentrated than the converged FLASH profile, which leads to a $t_0$ value that is higher by a few days. Using our default input physics increases the converged FLASH $\nimass$ by $\myapprox10\%$ (red lines in Figures~\ref{fig:CompareSimEjecta106}), mostly because of the inclusion of Coulomb corrections (see Table~\ref{tbl:Ni56 Sens Z0}). This also has the effect of decreasing $t_0$ by roughly a day, see Table~\ref{tbl:t0 Sens Z0}. We further calculate with V1D using our default input physics (black line in the top panel of Figure~\ref{fig:CompareSimEjecta106}, we use $f=0.05$) and we find that the $\nimass$ converged value deviates by $\myapprox0.5\%$ from the FLASH converged value, which is similar to the comparison of Section~\ref{sec:convergence}. 

All zero metallicity cases studied by \citet{Sim2010} are compared in Figure~\ref{fig:ComparePrevious} (right-pointing triangles connected with dashed black lines) to our default cases (solid black lines). The combination of Coulomb corrections omission and the scheme used by \citet{Sim2010} lead to the largest deviations from our default results, compared to other studies. Although \citet{Sim2010} results reduce somewhat the tension with the observed $t_0-\nimass$ relation, we used converged results and more accurate physical input to show that their results are not accurate enough.  

\subsection{Comparison to \citet{Moll2014}}
\label{sec:Moll 2014}

\citet{Moll2014} used KEPLER \citep[Lagrangian code;][]{Woosley2011} to calculate SCD. WD masses in the range of $0.8-1.1\,M_{\odot}$ with zero metallicity were considered. The hydrodynamical calculations contained initial cell sizes of $\Delta x_0\approx10-20\,\textrm{km}$\footnote{S. Woosley, private communication.} (varied along the WD; no convergence test is presented) and a $199$-isotope network (calculated \textit{in situ} without the use of post-processing). The detonations were ignited at the center of the WD by a $20$-cell hotspot (e.g., for the $M=0.8\,M_{\odot}$ case, there was a central region with a radius of $\myapprox450\,\textrm{km}$ with a temperature of $\myapprox2.1\times10^{9}\,\textrm{K}$ and then a roughly linear temperature gradient up to a radius of $\myapprox700\,\textrm{km}$ and a temperature of $\myapprox1.9\times10^{7}\,\textrm{K}$)\footnote{S. Woosley, private communication.}. The initial temperature outside the hotspot varied (e.g., for the $M=0.8\,M_{\odot}$ case, from $\myapprox1.9\times10^{7}\,\textrm{K}$ outside the hotspot to  $\myapprox1.15\times10^{7}\,\textrm{K}$ at the edge of the WD)\footnote{S. Woosley, private communication.}. The $t_0$ and $\nimass$ values obtained by \citet{Moll2014} are compared in Figure~\ref{fig:ComparePrevious} (upward-pointing triangles connected with dotted black lines) to our default cases (solid black lines). The agreement between the results is excellent, although, as we show for the $\MWD=0.8\,M_{\odot}$ case below, some differences exist between the results. 

\begin{figure}
\includegraphics[width=0.48\textwidth]{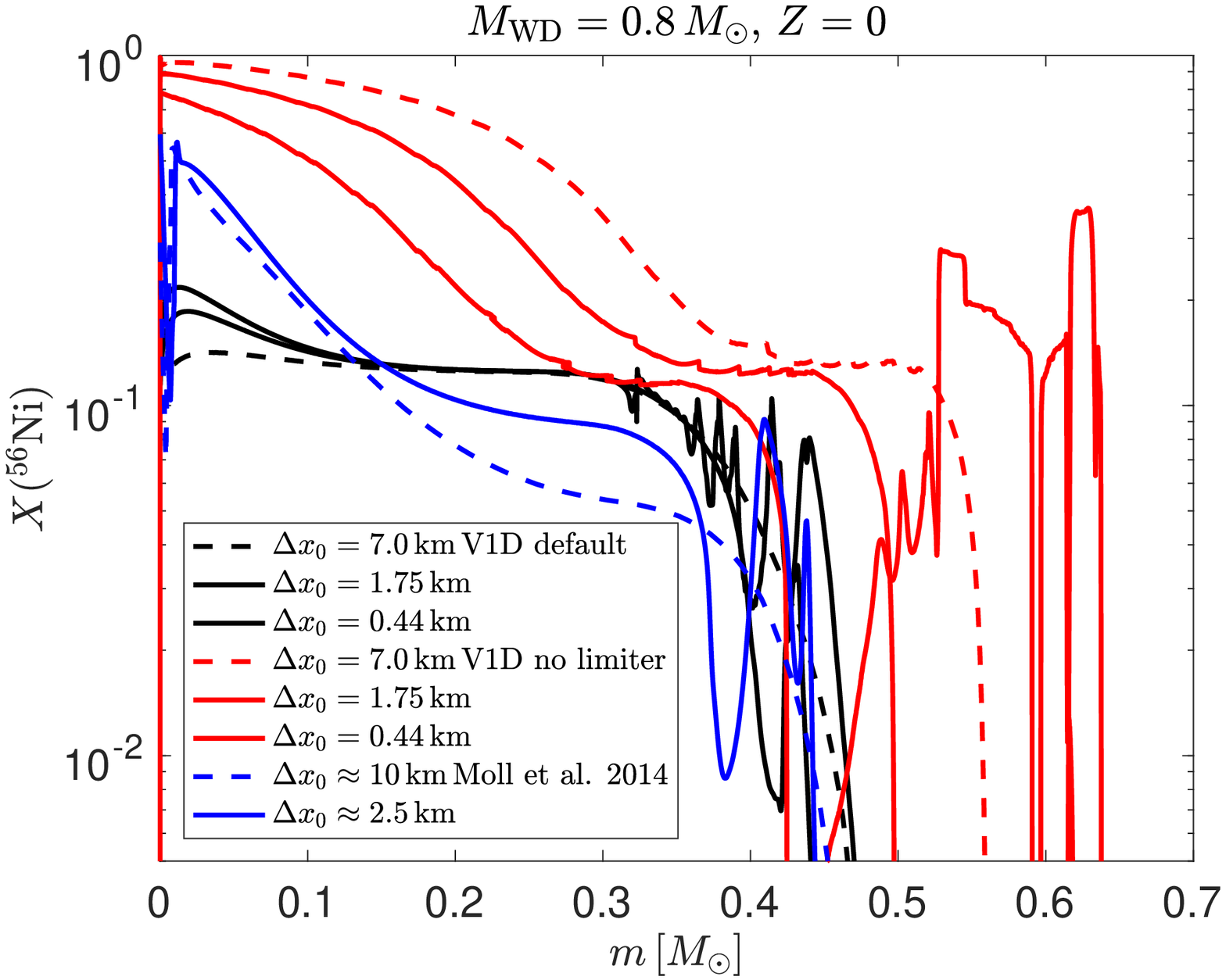}
\caption{Comparison to the $\MWD=0.8\,M_{\odot}$, $Z=0$ case of \citet{Moll2014}. We compare the $^{56}$Ni mass fraction distribution, $X(^{56}\rm{Ni})$, within the ejecta as a function of $m$, calculated with different resolutions (lowest resolution represented with a dashed line). Black lines: the default V1D results. Red lines: V1D results without using a burning limiter. Blue lines: the results of \citet{Moll2014} (with an additional high-resolution KEPLER calculation that was kindly provided to us by S. Woosley). The calculated \citet{Moll2014} $X(^{56}\rm{Ni})$ is higher (lower) than the default V1D results for $m\lesssim0.2\,M_{\odot}$ ($m\gtrsim0.2\,M_{\odot}$). Increasing the KEPLER resolution leads to better agreement with the V1D result for $m\gtrsim0.2\,M_{\odot}$. The convergence of the V1D results without a burning limiter is very slow, and the results deviate significantly from the converged results (for the typical resolutions used). Since the V1D results without a limiter and the KEPLER results seem to approach each other, we believe that a similar problem exists in the KEPLER calculations. The irregular $X(^{56}\rm{Ni})$ distribution at large $m$ values is related to the instability of the TNDW at low upstream densities (see Section~\ref{sec:convergence}). 
\label{fig:CompareWoosley08}}
\end{figure}

We compare in Figure~\ref{fig:CompareWoosley08} the $^{56}$Ni mass profiles within the ejecta of \citet[][dashed blue line]{Moll2014} to our default V1D calculations (black lines) for the $\MWD=0.8\,M_{\odot}$ case. As can be seen in the figure, the calculated \citet{Moll2014} $X(^{56}\rm{Ni})$ is higher (lower) than the default V1D results for $m\lesssim0.2\,M_{\odot}$ ($m\gtrsim0.2\,M_{\odot}$). Increasing the KEPLER resolution leads to better agreement with the V1D result for $m\gtrsim0.2\,M_{\odot}$\footnote{The high resolution KEPLER calculation was kindly provided to us by S. Woosley.}. Since the KEPLER code does not employ a burning limiter, we recalculate this case with V1D without a burning limiter (red lines). As can be seen in the figure, the convergence without a burning limiter is very slow (in principle, without a limiter, resolving the $\mysim1\,\textrm{cm}$ length scale of the TNDW is required), and the results deviate significantly from the converged results (for the typical resolutions used). Since the V1D results without a limiter and the KEPLER results seem to approach each other, we believe that a similar problem exists in the KEPLER calculations, although the different scheme adopted by KEPLER precludes direct comparison to V1D. The irregular $X(^{56}\rm{Ni})$ distribution at large $m$ values is related to the instability of the TNDW at low upstream densities (see Section~\ref{sec:convergence}). 

\begin{figure}
\includegraphics[width=0.48\textwidth]{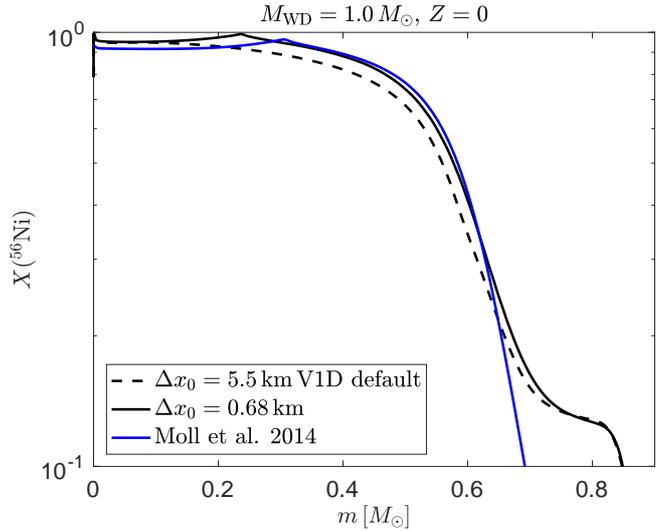}
\caption{Comparison to the $\MWD=1.0\,M_{\odot}$, $Z=0$ case of \citet{Moll2014}. We compare the $^{56}$Ni mass fraction distribution, $X(^{56}\rm{Ni})$, within the ejecta as a function of $m$. Black lines: the default V1D results for different resolutions (lowest resolution represented by a dashed line). Blue: the results of \citet{Moll2014}. The agreement between the results is very good.  
\label{fig:CompareWoosley10}}
\end{figure}

The $\MWD=1.0\,M_{\odot}$ case is compared in Figure~\ref{fig:CompareWoosley10}. Since in this case $X(^{56}\rm{Ni})$ almost reaches unity for a large fraction of the mass, the differences between the codes are much less pronounced, and the agreement is very good.  

\subsection{Comparison to \citet{Blondin2017}}
\label{sec:Blondin 2017}

\citet{Blondin2017} used a Lagrangian code to calculate SCD. WD masses in the range of $0.88-1.15\,M_{\odot}$ with solar metallicities were considered (no details regarding $T_{\rm{WD}}$ were provided). The hydrodynamical calculations contained a five-equation reaction scheme and post-processing with a $144$-isotope network \citep[see][for details]{Blondin2013}. The $t_0$ and $\nimass$ values obtained by \citet{Blondin2017} are compared in Figure~\ref{fig:ComparePrevious} (squares connected with dashed red lines) to our default cases (solid red lines). The deviations from our default results are quite large compared to those of other studies. At least part of this deviation can be explained by the lack of Coulomb corrections in the calculations of \citet{Blondin2017} (see Section~\ref{sec:Sim 2010} and Tables~\ref{tbl:Ni56 Sens Z1}-~\ref{tbl:t0 Sens Z1}). Because of the large effect of Coulomb corrections, we do not attempt here a more detailed comparison. 
 
\subsection{Comparison to \citet{Shen2018}}
\label{sec:Shen 2018}

\citet{Shen2018} used FLASH4.2.2 to calculate SCD. WD masses in the range of $0.8-1.1\,M_{\odot}$ with a few values for the metallicity and for the C/O were considered. The hydrodynamical calculations contained a $41$-isotope network and tracer particles for post-processing with a $205$-isotope network. They included a burning limiter to broaden the burning front over several cells, which is different than the one used here \citep{Kushnir2019} and by \citet{Kushnir2013}. In their implementation, changes in the temperature are limited within each hydrodynamical time-step, achieving broadened burning fronts and the ability to converge consistently. However, a few problems with this approach make the convergence properties uncertain \citep{Kushnir2019}. The detonations were ignited at the center of the WD by a hotspot with a radius of $400\,\textrm{km}$ that has a linear temperature gradient with a central temperature of $2\times10^{9}\,\textrm{K}$ and an outer temperature of $1.2\times10^{9}\,\textrm{K}$. The initial temperature outside the hotspot was set to $3\times10^{7}\,\textrm{K}$. The $t_0$ and $\nimass$ values obtained by \citet{Shen2018} for their zero metallicity and C/O$=50/50$ are compared in Figure~\ref{fig:ComparePrevious} (left-pointing triangles connected with dash-dotted black lines) to our default cases (solid black lines). There is reasonable agreement between the results, and in what follows we study in detail the reasons for the existing disagreement. 

We calculate the cases $\MWD=0.8,1\,M_{\odot}$ with solar metallicity and C/O$=50/50$ \citep[as defined by][$X(^{12}$C$)=X(^{16}$O$)=0.4945$, $X(^{22}$Ne$)=0.01$, $X(^{56}$Fe$)=0.001$ with $Y_{e}\approx0.4995$]{Shen2018}, which corresponds to $Z\approx0.7Z_{\odot}$ with our definition of solar metallicity (see Sections~\ref{sec:initial} and~\ref{sec:initial heavy}). We test both the ignition method of \citet{Shen2018} and our default ignition method (velocity gradient, see Section~\ref{sec:ignition}). We use both the input physics of \citet{Shen2018} and our default input physics. In order to match the input physics of \citet{Shen2018}, we use the \citet{Yakovlev1989} Coulomb corrections, we do not include the nuclear excitation correction to the EOS, and we use the \textsc{extended screening} option of {\sc MESA} for screening. Since the \textsc{extended screening} option does not respect a detailed balance, we do not use the ASE scheme for this comparison. We calculate with $204$ isotopes (the $205$-isotope list of \citet{Shen2018} without the extremely-short-lived $^{59}$Ge) and we do not correct the JINA total cross-sections of the reactions $^{12}$C+$^{16}$O and $^{16}$O+$^{16}$O. 

\begin{figure}
\includegraphics[width=0.48\textwidth]{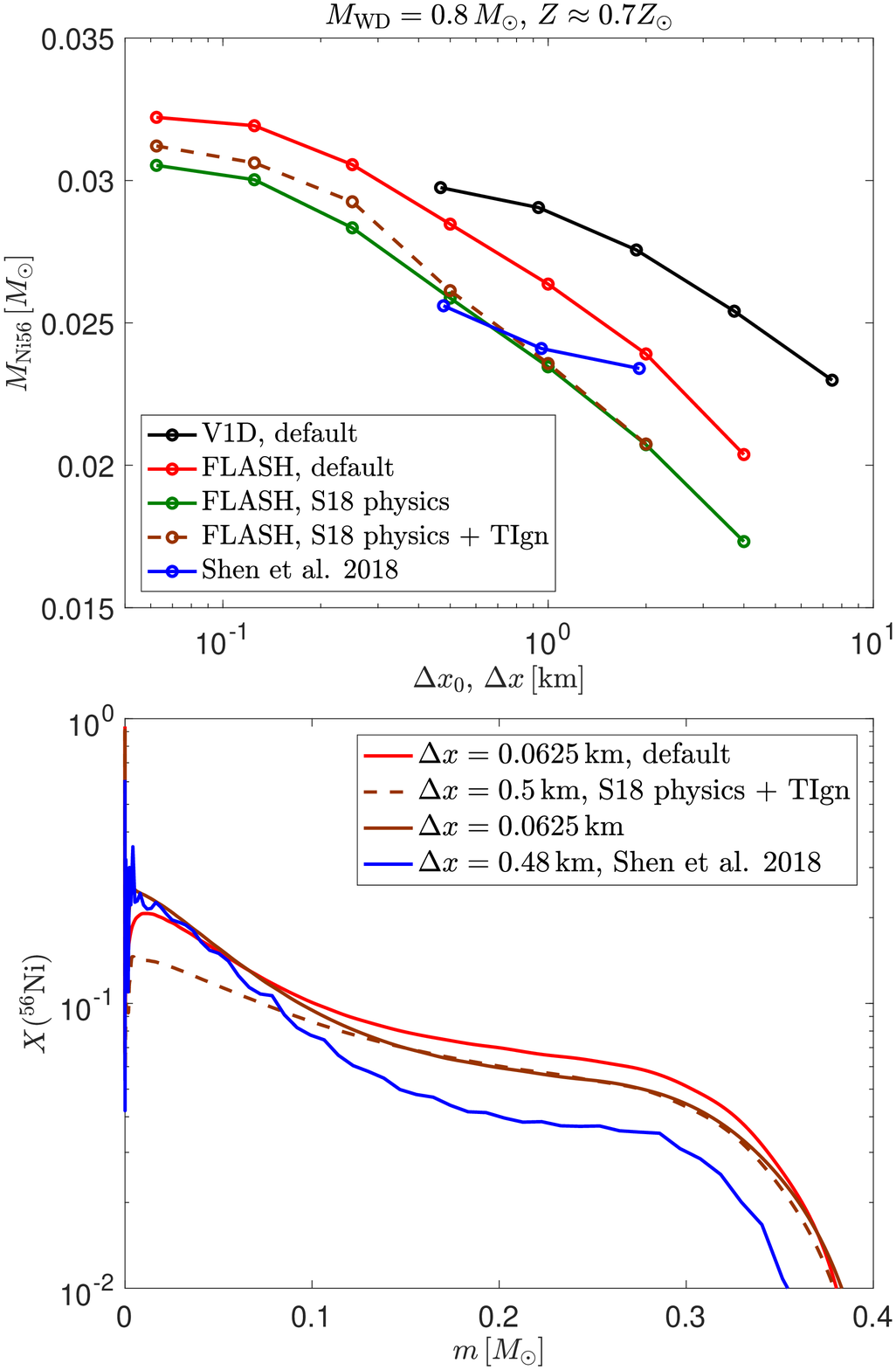}
\caption{Comparison to the $\MWD=0.8\,M_{\odot}$, $Z\approx0.7Z_{\odot}$ case of \citet{Shen2018}. Top panel: $\nimass$ as a function of the V1D initial (FLASH maximal) resolution, $\Delta x_0$ ($\Delta x$). Black (red) line: V1D (FLASH) results with our default input physics. Green line: FLASH results with \citet{Shen2018} input physics. Brown line: FLASH results with \citet{Shen2018} input physics and ignition method. Blue line: The result of \citet{Shen2018}. Bottom panel: The $^{56}$Ni mass fraction distribution, $X(^{56}\rm{Ni})$, within the ejecta, as a function of $m$. Red line: the converged ($\Delta x=0.0625\,\textrm{km}$) FLASH result with our default input physics. Brown lines: The $\Delta x=0.5\,\textrm{km}$ resolution (dashed line) and the converged ($\Delta x=0.0625\,\textrm{km}$, solid line) FLASH result with \citet{Shen2018} input physics and ignition method. Blue line: The result of \citet{Shen2018}. Our results in the top panel are similar to the results of \citet{Sim2010} for $\Delta x=0.48\,\textrm{km}$ and for $\Delta x=0.95\,\textrm{km}$, while there is $\myapprox12\%$ deviation for $\Delta x=1.91\,\textrm{km}$. However, this similarity is in part accidental, since the $^{56}$Ni mass profiles within the ejecta are somewhat different (compare the dashed brown and blue lines in the bottom panel). The results of \citet{Shen2018} do not converge, and the converged value we get is $\myapprox30\%$ higher than their highest resolution value (see also the solid brown line in the bottom panel). Using a velocity gradient ignition instead of a hotspot ignition slightly lowers the $\nimass$ for the highest resolutions. Using our default input physics increases the converged FLASH $\nimass$ by $\myapprox5\%$. The V1D $\nimass$ converged value is $\myapprox8\%$ lower than the FLASH converged value.
\label{fig:CompareShenEjecta08}}
\end{figure}

The values of $\nimass$ that we obtained for $\MWD=0.8\,M_{\odot}$ with the input physics of \citet{Shen2018} and with their ignition method (brown line, FLASH with $f=0.1$) are compared to the results of \citet{Shen2018} (blue line) in the top panel of Figure~\ref{fig:CompareShenEjecta08}\footnote{We thank K. Shen for providing us with the results of the lower resolution calculations.}. As can be seen in the figure, our results are similar to the results of \citet{Shen2018} for $\Delta x=0.48\,\textrm{km}$ and for $\Delta x=0.95\,\textrm{km}$, while there is $\myapprox12\%$ deviation for $\Delta x=1.91\,\textrm{km}$. However, this similarity is in part accidental, as the $^{56}$Ni mass profiles within the ejecta are somewhat different, see the bottom panel of Figure~\ref{fig:CompareShenEjecta08} (compare the dashed brown and blue lines). The results of \citet{Shen2018} do not converge, and the converged value we get is $\myapprox30\%$ higher than their highest resolution value (see also the solid brown line in the bottom panel of Figure~\ref{fig:CompareShenEjecta08}). Using a velocity gradient ignition (green line in the upper panel of Figure~\ref{fig:CompareShenEjecta08}) instead of a hotspot ignition slightly lowers $\nimass$ for the highest resolutions, since a smaller mass is being affected by the ignition region (which becomes smaller than the fixed $400\,\textrm{km}$ hotspot size). The velocity ignition also allows to ignite at low resolutions, where the hotspot ignition fails (for some fixed burning limiter). Using our default input physics increases the converged FLASH $\nimass$ by $\myapprox5\%$ (red lines in Figure~\ref{fig:CompareShenEjecta08}). We further calculate with V1D using our default input physics (black line in the top panel of Figure~\ref{fig:CompareShenEjecta08}, we use $f=0.1$) and find that the $\nimass$ converged value is lower by $\myapprox8\%$ than the FLASH converged value, which is similar to the results of the comparison in Section~\ref{sec:convergence}. 

\begin{figure}
\includegraphics[width=0.48\textwidth]{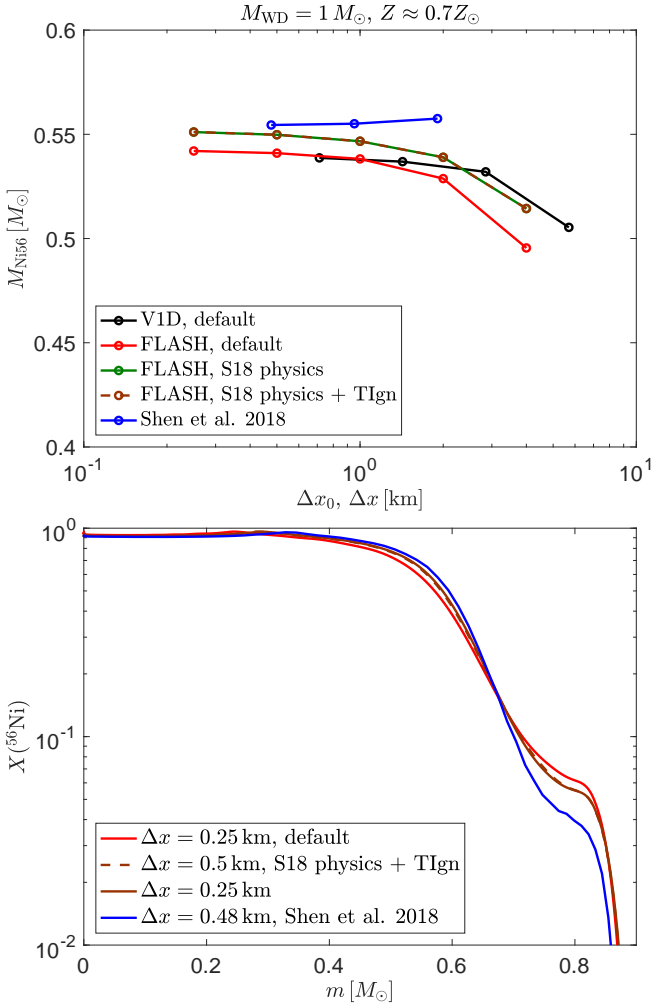}
\caption{Same as Figure~\ref{fig:CompareShenEjecta08} for $\MWD=1\,M_{\odot}$. Our results are similar to the results of \citet{Shen2018} and the results seem to converge to roughly the same value. This is also evident from comparing the $^{56}$Ni mass profiles within the ejecta in the bottom panel (compare the brown lines to the blue line). Using a velocity gradient ignition (green line in the top panel) instead of a hotspot ignition has no effect on the results. Using our default input physics decreases the converged FLASH $\nimass$ by $\myapprox1.5\%$ (red lines). The V1D $\nimass$ converged value deviates by $\myapprox0.5\%$ from the FLASH converged value.
\label{fig:CompareShenEjecta10}}
\end{figure}

The values of $\nimass$ that we obtained for $\MWD=1,M_{\odot}$ with the input physics of \citet{Shen2018} and with their ignition method (brown line, FLASH with $f=0.1$) are compared to the results of \citet{Shen2018} (blue line) in the top panel of Figure~\ref{fig:CompareShenEjecta10}. As can be seen in the figure, our results are similar to the results of \citet{Shen2018} and the results seem to converge to roughly the same value. This is also evident from comparing the $^{56}$Ni mass profiles within the ejecta in the bottom panel of Figure~\ref{fig:CompareShenEjecta10} (compare the brown lines to the blue line). Using a velocity gradient ignition (green line in the top panel of Figure~\ref{fig:CompareShenEjecta10}) instead of a hotspot ignition has no effect on the results. Using our default input physics decreases the converged FLASH $\nimass$ by $\myapprox1.5\%$ (red lines in Figure~\ref{fig:CompareShenEjecta10}). We further calculate with V1D using our default input physics (black line in the top panel of Figure~\ref{fig:CompareShenEjecta10}, we use $f=0.05$) and we find that the $\nimass$ converged value deviates by $\myapprox0.5\%$ from the FLASH converged value, which is similar to the results of comparison in Section~\ref{sec:convergence}.  

\subsection{Comparison to \citet{Miles2019}}
\label{sec:Miles 2019}

\citet{Miles2019} used FLASH to calculate SCD of solar metallicity $\MWD=0.8\,M_{\odot}$. The hydrodynamical calculations contained a $205$-isotope network and tracer particles for post-processing with the same network. The post-processing was done either in the usual way or with the reconstruction method \citep[see][for details]{Miles2019}. No burning limiter was included. The detonations were ignited at the center of the WD by a hotspot with a radius of $150\,\textrm{km}$ that has a linear temperature gradient with a central temperature of $1.98\times10^{9}\,\textrm{K}$ and an outer temperature of $3\times10^{7}\,\textrm{K}$, which is also the temperature outside the hotspot, out to a pressure of $10^{20}\,\textrm{dyne}\,\textrm{cm}^{-2}$. Below this pressure, they used $d\ln T/d \ln P=0.2$. They used a uniform composition of $X(^{12}$C$)=0.5$, $X(^{16}$O$)=0.4813$, $X(^{22}$Ne$)=0.014$, and $\Sigma_i X_i=0.047$ (with the definitions of Section~\ref{sec:initial heavy})\footnote{This is somewhat different from the values reported by \citet{Miles2019}, B. Miles, private communication.}, which corresponds to a solar metallicity with $Y_{e}\approx0.4993$ (see Section~\ref{sec:initial heavy} for details). The $\nimass$ value obtained by \citet{Miles2019} with their reconstruction method is compared in Figure~\ref{fig:ComparePrevious} (red filled circle) to our default cases (solid red line). There is reasonable agreement between the results, and in what follows we study in detail the reasons for the existing disagreement. 

We calculate the same case studied by \citet{Miles2019}, without assuming $d\ln T/d \ln P=0.2$ for the outer part of the profile. We test both the ignition method of \citet{Miles2019} and our default ignition method (a velocity gradient, see Section~\ref{sec:ignition}). We use both the input physics of \citet{Miles2019} and our default input physics. In order to match the input physics of \citet{Miles2019}, we use the \citet{Yakovlev1989} Coulomb corrections, we do not include the nuclear excitation correction to the EOS, and we use the \textsc{extended screening} option of {\sc MESA} for screening. Since the \textsc{extended screening} option does not respect a detailed balance, we do not use the ASE scheme for this comparison. We calculate with the same $205$-isotope list of \citet{Miles2019} and we do not correct the JINA total cross-sections of the reactions $^{12}$C+$^{16}$O and $^{16}$O+$^{16}$O. 

\begin{figure}
\includegraphics[width=0.48\textwidth]{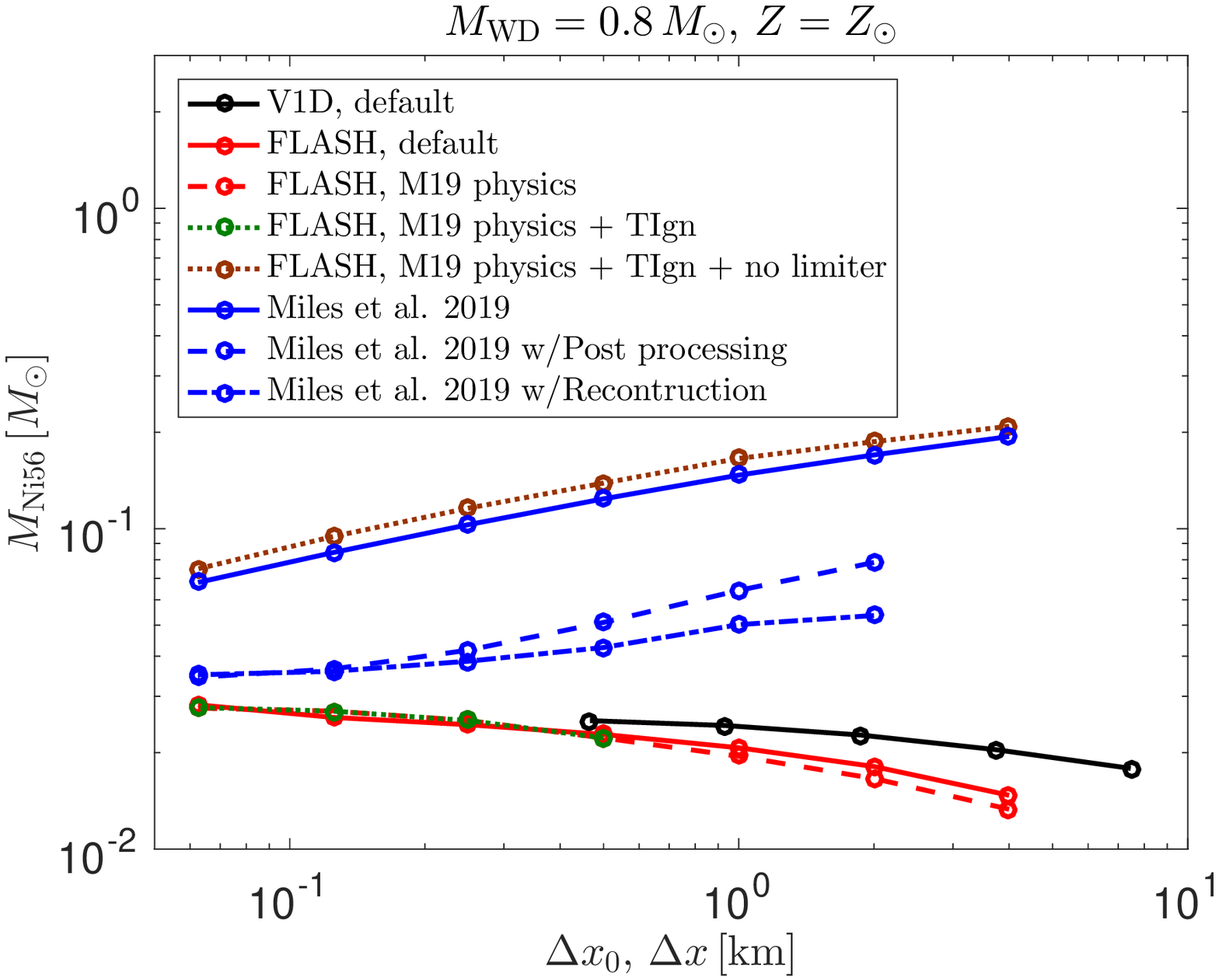}
\caption{Comparison to the $\MWD=0.8\,M_{\odot}$, $Z=Z_{\odot}$ case of \citet{Miles2019}. We compare the $\nimass$ as a function of the V1D initial (FLASH maximal) resolution, $\Delta x_0$ ($\Delta x$). Black (solid red) line: V1D (FLASH) results with our default input physics. Dashed red line: FLASH results with \citet{Miles2019} input physics. Green line: FLASH results with the input physics and the ignition method of \citet{Miles2019}. Brown line: FLASH calculations without a burning limiter, with the input physics and the ignition method of \citet{Miles2019}. Blue lines: The result of \citet{Miles2019}, with solid, dashed and dash-dotted lines corresponding to no post-processing, regular post-processing, and post-processing with reconstruction, respectively. The FLASH results without a burning limiter are similar to results of \citet{Miles2019} when no post-processing was applied. The FLASH results with a burning limiter converge to a value that is $\myapprox25\%$ lower than the converged value obtained from the usual post-processing  and from the reconstruction method. Using a velocity gradient ignition instead of a hotspot ignition has a minimal effect. Using our default input physics has a minimal effect as well. The converged $\nimass$ obtained with V1D using our default input physics and ignition method is $\myapprox11\%$ lower than the FLASH converged value.
\label{fig:CompareMiles}}
\end{figure}

In Figure~\ref{fig:CompareMiles}, we compare the value of $\nimass$ that we obtained with the input physics of \citet{Miles2019} and with their ignition method (green line, with $f=0.1$, brown line without burning limiter) to the results of \citet{Miles2019} (blue lines). As can be seen in the figure, the results of the calculations without the burning limiter are similar to results of \citet{Miles2019} when no post-processing was applied (solid blue line). Our FLASH results with a burning limiter converge to a value that is $\myapprox25\%$ lower than the converged value obtained from the usual post-processing (dashed blue line) and from the reconstruction method (dotted blue line). We believe that the reason for this deviation is that the underlying simulation for the post-processing did not contain a burning limiter, such that the input for the post-processing procedure is far from the converged input. It is not clear how the post-processing procedures can completely take this into account. Using a velocity gradient ignition (dashed red line) instead of a hotspot ignition has a minimal effect, since the size of the hotspot, $150\,\textrm{km}$, is quite small. However, high resolution, $\Delta x\lesssim0.5\,\textrm{km}$, is required to ignite such a small hotspot size (for $f=0.1$). Using our default input physics (solid red line) has a minimal effect as well. We further calculate with V1D using our default input physics and ignition method (black line, we use $f=0.1$) and find that the $\nimass$ converged value is lower by $\myapprox11\%$ than our FLASH converged value, which is similar to the results of the comparison in Section~\ref{sec:convergence}.

\subsection{Comparison to \citet{Bravo2019}}
\label{sec:Bravo 2019}

\citet{Bravo2019} used a Lagrangian PPM code \citep[based on][]{Colella1984,Colella1985} to calculate SCD. WD masses in the range of $0.88-1.15\,M_{\odot}$ and a few metallicity values were considered (with $T_{\rm{WD},9}=0.1$). The hydrodynamical calculations of the $\MWD=0.88(1.1)\,M_{\odot}$ case contained initial cell sizes of $\Delta x_0\approx7-25(4-15)\,\textrm{km}$ (which varied along the WD; no convergence test is presented). The nuclear network was calculated \textit{in situ} without the use of post-processing. The number of isotopes in each cell changed adaptively, and could reach $722$ isotopes. No burning limiter was included. The TNDW was ignited at the center of the WD, but details regarding the ignition method are not given. The $t_0$ and $\nimass$ values obtained by \citet{Bravo2019} for their roughly zero metallicity ($Z\sim10^{-2}Z_{\odot}$) and C/O$=50/50$ calculations are compared in Figure~\ref{fig:ComparePrevious} (downward-pointing triangles connected with dotted blue lines) to our default cases (solid black lines). There is reasonable agreement between the results, and in what follows we study in detail the reasons for the existing disagreement.

We calculate with V1D the cases $\MWD=0.88,1.1\,M_{\odot}$ with zero metallicity and C/O$=50/50$. We use our default ignition method (velocity gradient, see Section~\ref{sec:ignition}). The differences between our default input physics and the default physics of \citet{Bravo2019} include a slightly different EOS, the inclusion of the nuclear excitation correction to the EOS, a different prescription of the reaction rate screening, and a different cross-section for the reaction $^{16}$O+$^{16}$O. Furthermore, \citet{Bravo2019} probably used a more extensive isotopes-list than our $178$-isotope list for most of the cells. Since all these differences have a small effect on the results (see Sections~\ref{sec:uncertainty},~\ref{sec:reaction rate sensitivity} and the previous sub-sections of this section), we use our default input physics for this comparison.

\begin{figure}
\includegraphics[width=0.48\textwidth]{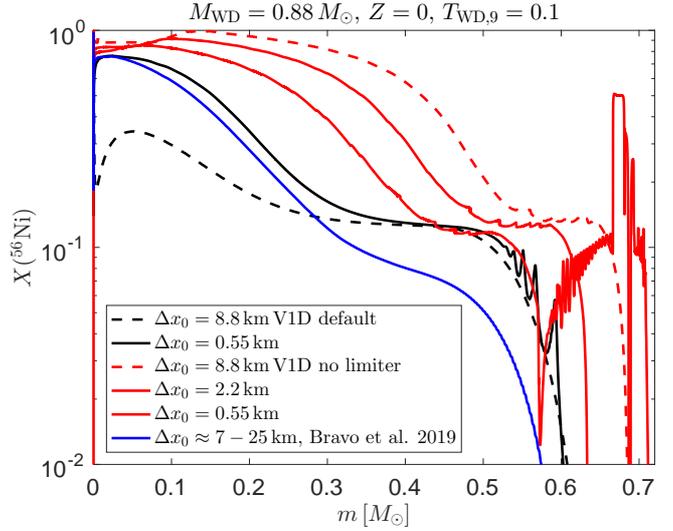}
\caption{Comparison to the $\MWD=0.88\,M_{\odot}$, $Z=0$ case of \citet{Bravo2019}. We compare the $^{56}$Ni mass fraction distribution, $X(^{56}\rm{Ni})$, within the ejecta, as a function of $m$, calculated with different resolutions (lowest resolution represented with a dashed line). Black lines: the default V1D results. Red lines: V1D results without using a burning limiter. Blue line: The result of \citet{Bravo2019}. The convergence of the V1D results without a burning limiter is very slow, and $X(^{56}\rm{Ni})$ deviate significantly from the converged $X(^{56}\rm{Ni})$ (for the typical resolutions used). We believe that the results of \citet{Bravo2019} are not converged, although their different scheme precludes a direct comparison to V1D. The irregular $X(^{56}\rm{Ni})$ distribution at large $m$ values is related to the instability of the TNDW at low upstream densities (see Section~\ref{sec:convergence}). 
\label{fig:CompareBravo088}}
\end{figure}

We compare in Figure~\ref{fig:CompareBravo088} the $^{56}$Ni mass profiles within the ejecta of \citet[][blue line]{Bravo2019} to our default default V1D calculations (black lines) for the $\MWD=0.88\,M_{\odot}$ case. As can be seen in the figure, the calculated \citet{Bravo2019} $X(^{56}\rm{Ni})$ is lower than the default V1D results, most pronouncedly for $m\gtrsim0.2\,M_{\odot}$. Since the code used by \citet{Bravo2019} does not employ a burning limiter, we recalculate this case with V1D without a burning limiter (red lines). As can be seen in the figure, the convergence without a burning limiter is very slow (in principle, without a limiter, resolving of the $\mysim1\,\textrm{cm}$ length scale of the TNDW is required), and the $^{56}$Ni mass profile deviates significantly from the converged $^{56}$Ni mass profile (for the typical resolutions used). We believe that the results of \citet{Bravo2019} are not converged, although their different scheme precludes a direct comparison to V1D. The irregular $X(^{56}\rm{Ni})$ distribution at large $m$ values is related to the instability of the TNDW at low upstream densities (see Section~\ref{sec:convergence}). 

\begin{figure}
\includegraphics[width=0.48\textwidth]{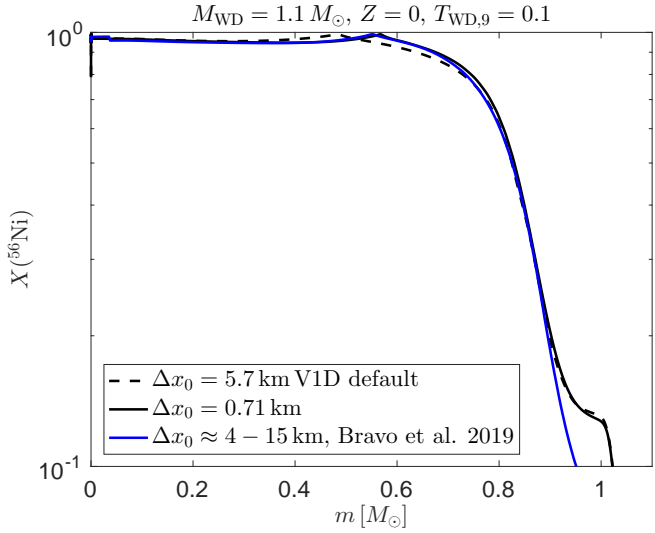}
\caption{Comparison to the $\MWD=1.1\,M_{\odot}$, $Z=0$ case of \citet{Bravo2019}. We compare the $^{56}$Ni mass fraction distribution, $X(^{56}\rm{Ni})$, within the ejecta, as a function of $m$. Black lines: The default V1D results for different resolutions (lowest resolution represented by a dashed line). Blue line: The result of \citet{Bravo2019}. The agreement between the results is very good.  
\label{fig:CompareBravo11}}
\end{figure}

The $\MWD=1.1\,M_{\odot}$ case is compared in Figure~\ref{fig:CompareBravo11}. Since in this case $X(^{56}\rm{Ni})$ is almost reaches unity for a large fraction of the mass, the differences between the codes are much less pronounced, and the agreement is very good.  

\section{Summary and discussion}
\label{sec:discussion}

We have used V1D and FLASH, modified to include an accurate and efficient burning scheme (Section~\ref{sec:setup}), to perform 1D calculations of SCD, focusing on the recently compiled $t_0-\nimass$ relation, where $t_0$ (the $\gamma$-rays escape time from the ejecta, which is measured to an accuracy of a few percent) is positively correlated with $\nimass$ (the $^{56}$Ni mass synthesized in the explosion). The advantage of comparing to this relation is that it bypasses the need for radiation transfer calculations. We showed in Section~\ref{sec:results} that the calculated $\nimass$ and $t_0$ converge to an accuracy better than a few percent. The converged results of these calculations are presented in Figure~\ref{fig:Nit0}, which is the main result of this work. As can be seen in the figure, there is a clear tension between the predictions of SCD and the observed $t_0-\nimass$ relation. SCD predicts an anti-correlation between $t_0$ and $\nimass$, with $t_0\approx30\,\textrm{day}$ for luminous ($\nimass\gtrsim0.5\,M_{\odot}$) SNe Ia, while the observed $t_0$ is in the range of $35-45\,\textrm{day}$. We showed that various uncertainties related to the physical processes and to the initial profiles of the WD are unlikely to resolve the tension with the observations (Section~\ref{sec:WD sensitivity}), while they can reduce the agreement with the observations for low-luminosity SNe Ia. We calibrated in Section~\ref{sec:small network} a $69$-isotope network, for which the $t_0-\nimass$ relation is accurately calculated. We then used this reduced network to perform in Section~\ref{sec:reaction rate sensitivity} a sensitivity check of our results to uncertainties in the reaction rate values. We found that the tension between the predictions of this model and the observed $t_0-\nimass$ relation is much larger than the uncertainty related to reaction rates, which, again, can decrease the agreement with the observations for low-luminosity SNe Ia.

In Section~\ref{sec:comparison}, we compared our results to previous studies of the problem, performed with less accurate numerical schemes. We showed that the general $\nimass-\MWD$ and $t_0-\nimass$ relations (Figure~\ref{fig:ComparePrevious}) are reproduced in all previous works \citep[except for the results of][which are systematically different from all other works, see Section~\ref{sec:Sim 2010}]{Sim2010}. Specifically, the tension with the observed $t_0-\nimass$ relation exists in all previous studies. 

We studied the effect of the initial composition on the $t_0-\nimass$ relation in Section~\ref{sec:C/O ratio}, where we calculated with an initial composition of C/O$\myapprox30/70$, which corresponds to the smallest $^{12}$C fraction suggested by evolutionary models of WDs \citep[][our default initial composition is C/O$=50/50$]{Renedo2010,Lauffer2018}. We found that $t_0$ increases for C/O$\myapprox30/70$, with respect to the C/O$=50/50$ case, for all $\nimass$ values (see Figure~\ref{fig:Nit0_C3070}). While the increase was insufficient to explain the observations of luminous SNe Ia, it is possible that even heavier initial composition would bring the calculated $t_0$ into better agreement with the observations. Such a heavier initial composition is indeed expected for $\MWD\gtrsim1.1\,M_{\odot}$ \citep{Lauffer2018}, however, the exact $\MWD$ for the transition as well as the exact initial composition (for all WD masses) are quite uncertain. We intend to study in subsequent work whether there exist some initial compositions that can reproduce the $t_0-\nimass$ relation.

The tension between the predictions of SCD and the observed $t_0-\nimass$ relation necessitates modifications to this simple model. A valid question is whether the modifications of DDM are sufficient to resolve the tension. Although the nucleosynthesis and the energy release within the thin helium layer are unlikely to affect neither $\nimass$ nor $t_0$, it is not clear what would be the effect of the compression wave that propagates in the CO core prior to ignition and the off-centre ignition. In order to study these effects, multi-D simulations with an accuracy of a few percent are required, which are not available today \citep[for example, the DDM calculation of][has only $\Delta x=4\,\textrm{km}$ and no burning limiter]{Townsley2019}. Nevertheless, we find it unlikely that these effects could significantly decrease the tension with the observed $t_0-\nimass$ relation. The compression wave only slightly increases the density prior to ignition, and probably leads to $^{56}$Ni synthesis further out in the WD and a decrease of $t_0$ (we verified this effect with 1D models that will be reported elsewhere). The off-centre ignition would lead to a scatter around the 1D $t_0-\nimass$ relation, but the tension with the observations is systematic, where the prediction is $t_0\approx30\,\textrm{day}$ for luminous SNe Ia, while the observations are in the range of $35-45\,\textrm{day}$. Another possibility is that the ejecta interacts with a companion (that exists in some versions of DDM), which would increase $t_0$. However, this process should have an observable effect only in a fraction of the viewing angles.

There are some reasonable initial compositions and reaction rate values for which SCD successfully explains the low-luminosity part of the $t_0-\nimass$ relation. However, DDM seems to be in conflict with the $^{56}$Ni mass-weighted line-of-sight velocity distribution for a large fraction of these events, as measured from nebular spectra observations \citep{Dong2015,Dong2018,Vallely2020}. Specifically, the $^{56}$Ni velocity distribution is either double-peaked or highly shifted, which is difficult to reconcile with DDM. These studies and the current work raise the possibility that DDM is unable to explain consistently any part of the SNe Ia luminosity range.  

It was already established by \citet{Wygoda2019a} that Chandrasekhar-mass models are unable to explain the $t_0-\nimass$ relation for low-luminosity SNe Ia. Taken together with the tension of sub-Chandrasekhar mass models to explain the $t_0-\nimass$ relation for high-luminosity SNe Ia presented here raises the question whether any model can consistently explain the full range of the $t_0-\nimass$ relation. Specifically, both Chandrasekhar-mass and the sub-Chandrasekhar mass models predict an anti-correlation between $t_0$ and $\nimass$. The direct-collision model \citep{Kushnir2013} has already showed some hints of success in explaining the entire $t_0-\nimass$ relation \citep{Wygoda2019a}. However, in order to establish this success, multi-D simulations with an accuracy of a few percent are required, which are not available today. We believe that our new scheme, together with accurate small reaction networks (similar to the $69$-isotope network that we calibrated), may allow such calculations in the near future. 


\section*{Acknowledgements}
We thank Stan Woosley, St\'ephane Blondin, Ken Shen, Eduardo Bravo and Stuart Sim for sharing their ejecta profiles with us and for useful discussions. We thank Boaz Katz, Subo Dong, Dean Townsley and Borxton Miles for useful discussions. DK is supported by the Israel Atomic Energy Commission -- The Council for Higher Education -- Pazi Foundation -- and by a research grant from The Abramson Family Center for Young Scientists.  

\section*{Data availability}
All ejecta profiles used to derive the results in this paper (except for the results in Section~\ref{sec:reaction rate sensitivity}), as well as the bolometric light curves from Section~\ref{sec:t0 accuracy}, are publicly available through \url{https://www.dropbox.com/s/3kd8te2yimdxotm/CIWD.tar.gz?dl=0}.






\begin{appendix}

\section{Input physics}
\label{sec:input}

Our input physics, which we briefly summarize below, are the ones used by \citet{KK2019}. A detailed description can be found in \citep{Kushnir2019}.

The nuclear masses are taken from the file \textsc{winvn\_v2.0.dat}, which is available through the JINA reaclib data base\footnote{http://jinaweb.org/reaclib/db/} \citep[JINA,][]{Cyburt2010}. For the partition functions, $w_{i}(T)$, we use the fit of \citet{Kushnir2019} for the values that are provided in the file \textsc{winvn\_v2.0.dat} over some specified temperature grid. The forward reaction rates are taken from JINA (the default library of 2017 October 20). All strong reactions that connect between isotopes from the list are included. Inverse reaction rates are determined according to a detailed balance. Enhancement of the reaction rates due to screening corrections is described at the end of this section. We further normalize all the channels of the $^{12}$C+$^{16}$O and $^{16}$O+$^{16}$O reactions such that the total cross-sections are identical to the ones provided by \citet{CF88}, while keeping the branching ratios provided by JINA. Unless stated otherwise, we ignore weak reactions and thermal neutrino emission. 

The EOS is composed of contributions from electron--positron plasma, radiation, ideal gas for the nuclei, ion--ion Coulomb corrections and nuclear level excitations. We use the EOS provided by {\sc MESA} for the electron--positron plasma, for the ideal gas part of the nuclei, for the radiation and for the Coulomb corrections (but based on \citet{Chabrier1998} and not on \citet{Yakovlev1989}, see below). The electron--positron part is based on the \textit{Helmholtz} EOS \citep{Timmes00}, which is a table interpolation of the Helmholtz free energy as calculated by the Timmes EOS\footnote{http://cococubed.asu.edu/} \citep{Timmes1999} over a density-temperature grid with $20$ points per decade. This is different from \citet{Kushnir2019}, where the Timmes EOS was used for the electron--positron plasma, since the \textit{Helmholtz} EOS is more efficient and because the internal inconsistency of the \textit{Helmholtz} EOS \citep[see][for details]{Kushnir2019} is small enough within the regions of the parameter space studied here. We further include the nuclear level excitation energy of the ions, by using the $w_{i}(T)$ from above.

We assume the Coulomb corrections to the chemical potential of each ion are given by $\mu_{i}^{C}=k_{B}Tf(\Gamma_{i})$ and are independent of the other ions \citep[linear mixing rule (LMR),][]{Hansen1977}, where $k_{B}$ is Boltzmann's constant, $\Gamma_{i}=Z_{i}^{5/3}\Gamma_{e}$ is the ion coupling parameter, where $Z_i$ is the proton number, and $\Gamma_{e}\approx(4\upi\rho N_{A} Y_{e}/3)^{1/3}e^{2}/k_{B}T$ is the electron coupling parameter, where $N_{A}$ is Avogadro's number and $Y_e\approx\sum_i X_i Z_i/A_i$ is the electron fraction. We use the three-parameter fit of \citet{Chabrier1998} for $f(\Gamma)$.  Following \citet[][]{Khokhlov88}, we approximate the LMR correction to the EOS by $f(\Gamma)$ for a `mean' nucleus $\Gamma=\bar{Z}^{5/3}\Gamma_{e}$, where
\begin{eqnarray}
\bar{Z}=\frac{\sum_i Y_i Z_i}{\sum_i Y_i}.
\end{eqnarray}
The screening factor for a thermonuclear reaction with reactants $i=1,..,N$ and charges $Z_{i}$ is determined from a detailed balance \citep{KushnirScreen}:
\begin{eqnarray}\label{eq:NSE screening}
\exp\left(\frac{\sum_{i=1}^{N}\mu_{i}^{C}-\mu_{j}^{C}}{k_{B}T}\right),
\end{eqnarray}
where isotope $j$ has a charge $Z_{j}=\sum_{i=1}^{N}Z_{i}$ \citep[same as equation~(15) of][for the case of $N=2$]{Dewitt1973}.  

\section{Some more results for $Z\ne Z_{\odot}$}
\label{sec:more results}

\begin{figure}
\includegraphics[width=0.48\textwidth]{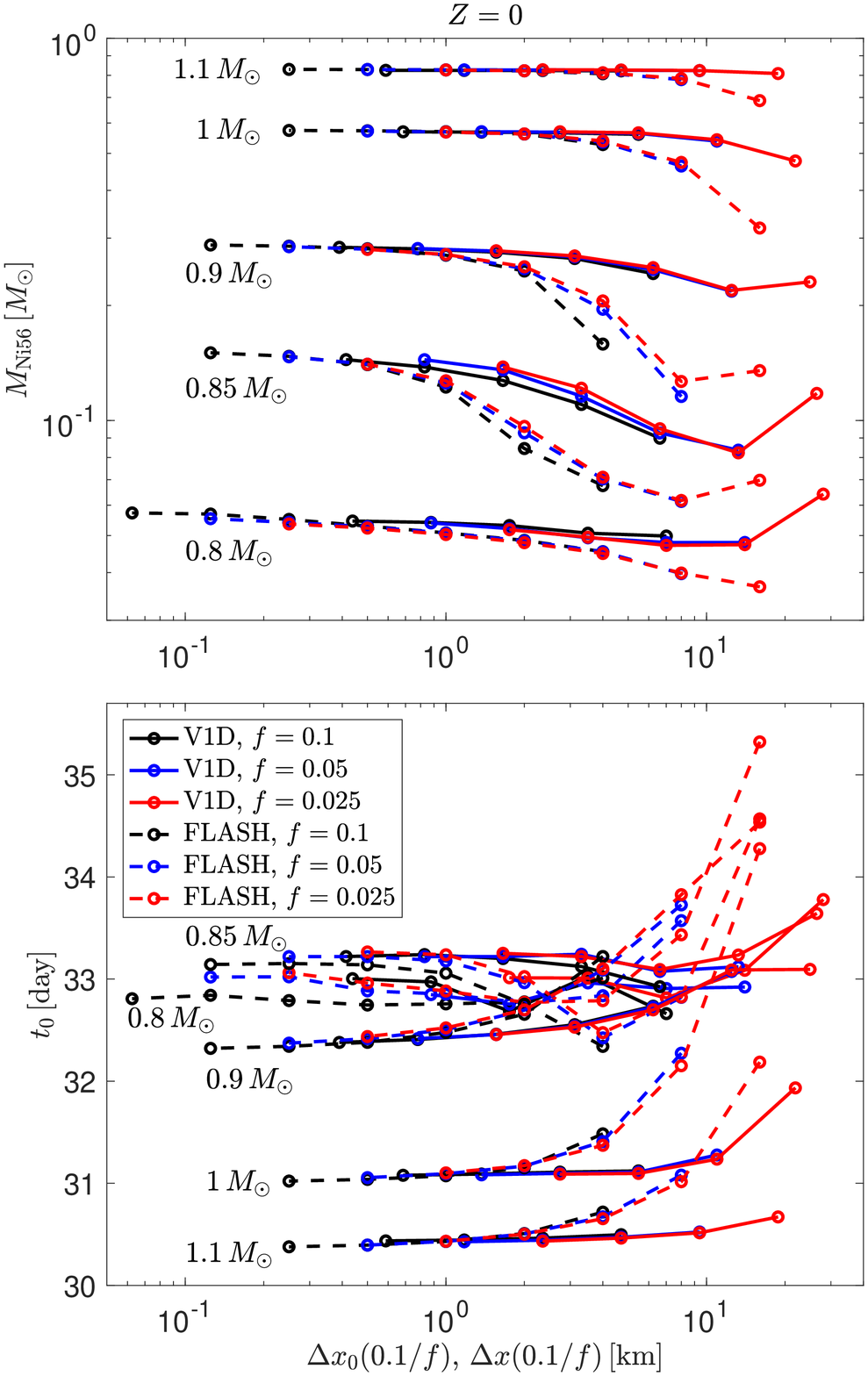}
\caption{Same as Figure~\ref{fig:Ni-t0_Convergence_Z1} for $Z=0$.
\label{fig:Ni-t0_Convergence_Z0}}
\end{figure}

In this appendix, we provide some more results (Figure~\ref{fig:Ni-t0_Convergence_Z0} and Tables~\ref{tbl:Ni56 Conv Z0}-\ref{tbl:Ni56 Sens Z0 FLASH}) of the calculations with $Z=0,0.5,2\,Z_{\odot}$. 

\begin{table*}
\begin{minipage}{160mm}
\caption{Same as Table~\ref{tbl:Ni56 Conv Z1} for $Z=0$.}
\begin{tabular}{|c||c||c||c||c||c||c|}
\hline
$\MWD\,[M_{\odot}]$  &$\Delta x_0,\,\Delta x\,[\textrm{km}]$ & $f$ & $\nimass\,[M_{\odot}]$ & $t_0\,[\textrm{day}]$  & $t_0^{\gamma\rm{RT}}\,[\textrm{day}]$ & $t_0^{\rm{RT}}\,[\textrm{day}]$ \\ \hline
0.8	&	$\Delta x_0$: 0.44	&	0.1	&	0.055	&	33.0	&	32.9	&	32	\\
     	&		&		&	(0.81\%)	&	(0.10\%)	&		&		\\
     	&	$\Delta x$: 0.0625	&	0.1	&	0.057	&	32.8	&	32.3	&	32	\\
     	&		&		&	(0.65\%)	&	(0.10\%)	&		&		\\
     0.85	&	$\Delta x_0$: 0.41	&	0.1	&	0.144	&	33.2	&	32.8	&	32	\\
     	&		&		&	(4.46\%)	&	(0.06\%)	&		&		\\
     	&	$\Delta x$: 0.125	&	0.1	&	0.150	&	33.1	&	32.4	&	32	\\
     	&		&		&	(2.13\%)	&	(0.03\%)	&		&		\\
     0.9	&	$\Delta x_0$: 0.39	&	0.1	&	0.284	&	32.4	&	31.9	&	32	\\
     	&		&		&	(0.91\%)	&	(0.09\%)	&		&		\\
     	&	$\Delta x$: 0.125	&	0.1	&	0.288	&	32.3	&	32.0	&	31	\\
     	&		&		&	(0.71\%)	&	(0.06\%)	&		&		\\
     1	&	$\Delta x_0$: 0.68	&	0.05	&	0.570	&	31.1	&	30.4	&	31	\\
     	&		&		&	(0.29\%)	&	(0.05\%)	&		&		\\
     	&	$\Delta x$: 0.25	&	0.1	&	0.574	&	31.0	&	30.2	&	31	\\
     	&		&		&	(0.21\%)	&	(0.05\%)	&		&		\\
     1.1	&	$\Delta x_0$: 0.59	&	0.05	&	0.827	&	30.4	&	29.5	&	31	\\
     	&		&		&	(0.07\%)	&	(0.05\%)	&		&		\\
     	&	$\Delta x$: 0.25	&	0.1	&	0.829	&	30.4	&	29.8	&	31	\\
     	&		&		&	(0.10\%)	&	(0.05\%)	&		&		\\
\hline
\end{tabular}
\centering
\label{tbl:Ni56 Conv Z0}
\end{minipage}
\end{table*}

\begin{table*}
\begin{minipage}{160mm}
\caption{The converged $Z=0.5Z_{\odot}$ values of $\nimass$ (4th column) and $t_0$ (5th column), as a function of $\MWD$ (1st column). We use the required V1D initial cell size, $\Delta x_{0}$ (2nd column), and $f$ values (3rd column) for the convergence of the $Z=0$ and $Z=Z_{\odot}$ results. The $t_0$ values estimated with the MC $\gamma$-ray transport (full radiation transfer) calculations are given in the 6th (7th) column.}
\begin{tabular}{|c||c||c||c||c||c||c|}
\hline
$\MWD\,[M_{\odot}]$  &$\Delta x_0\,[\textrm{km}]$ & $f$ & $\nimass\,[M_{\odot}]$ & $t_0\,[\textrm{day}]$ & $t_0^{\gamma\rm{RT}}\,[\textrm{day}]$& $t_0^{\rm{RT}}\,[\textrm{day}]$ \\ \hline
0.8	&	0.44	&	0.1	&	0.036	&	34.3	&	34.3	&	33	\\
     0.85	&	0.41	&	0.1	&	0.129	&	34.0	&	33.7	&	33	\\
     0.9	&	0.39	&	0.1	&	0.267	&	32.8	&	32.6	&	32	\\
     1	&	0.68	&	0.05	&	0.552	&	31.2	&	30.6	&	31	\\
     1.1	&	0.59	&	0.05	&	0.810	&	30.5	&	29.7	&	31	\\
\hline
\end{tabular}
\centering
\label{tbl:Ni56 Conv Z05}
\end{minipage}
\end{table*}

\begin{table*}
\begin{minipage}{160mm}
\caption{Same as Table~\ref{tbl:Ni56 Conv Z05} for $Z=2Z_{\odot}$.}
\begin{tabular}{|c||c||c||c||c||c||c|}
\hline
$\MWD\,[M_{\odot}]$  &$\Delta x_0\,[\textrm{km}]$ & $f$ & $\nimass\,[M_{\odot}]$ & $t_0\,[\textrm{day}]$ & $t_0^{\gamma\rm{RT}}\,[\textrm{day}]$& $t_0^{\rm{RT}}\,[\textrm{day}]$ \\ \hline
0.8	&	0.43	&	0.1	&	0.018	&	36.1	&	36.1	&	35	\\
     0.85	&	0.41	&	0.1	&	0.114	&	34.6	&	34.4	&	34	\\
     0.9	&	0.39	&	0.1	&	0.250	&	33.0	&	32.5	&	32	\\
     1	&	0.68	&	0.05	&	0.519	&	31.2	&	30.5	&	31	\\
     1.1	&	0.58	&	0.05	&	0.762	&	30.4	&	29.8	&	30	\\
\hline
\end{tabular}
\centering
\label{tbl:Ni56 Conv Z2}
\end{minipage}
\end{table*}

\begin{table*}
\begin{minipage}{160mm}
\caption{Same as Table~\ref{tbl:Ni56 Sens Z1} for $Z=0$.}
\begin{tabular}{|c||c||c||c||c||c||c||c||c||c||c|}
\hline
$\MWD\,[M_{\odot}]$  & $\Delta x_0\,[\textrm{km}]$ & Reference & NSE6  & NSE7 & w/o ASE & w weak & w thermal $\nu$ & w/o Coul. & w/o ex. & w/o screen \\ \hline
0.8	&	1.75	&	0.0531	&	0.0531	&	0.0531	&	0.0531	&	0.0531	&	0.0531	&	0.0402	&	0.0534	&	0.0506	\\
	&		&	(2.70\%)	&	(0.009\%)	&	(0.001\%)	&	(0.007\%)	&	(0.047\%)	&	(0.000\%)	&	(27.7\%)	&	(0.61\%)	&	(4.9\%)	\\
0.85	&	0.83	&	0.1380	&	0.1380	&	0.1380	&	0.1380	&	0.1380	&	0.1380	&	0.0766	&	0.1400	&	0.1171	\\
	&		&	(4.46\%)	&	(0.004\%)	&	(0.002\%)	&	(0.021\%)	&	(0.037\%)	&	(0.004\%)	&	(57.3\%)	&	(1.47\%)	&	(16.4\%)	\\
0.9	&	1.56	&	0.2756	&	0.2756	&	0.2756	&	0.2756	&	0.2757	&	0.2756	&	0.1985	&	0.2773	&	0.2522	\\
	&		&	(2.93\%)	&	(0.000\%)	&	(0.000\%)	&	(0.004\%)	&	(0.039\%)	&	(0.005\%)	&	(32.5\%)	&	(0.60\%)	&	(8.9\%)	\\
1	&	2.74	&	0.5638	&	0.5638	&	0.5638	&	0.5638	&	0.5639	&	0.5638	&	0.4821	&	0.5649	&	0.5388	\\
	&		&	(1.03\%)	&	(0.001\%)	&	(0.001\%)	&	(0.004\%)	&	(0.024\%)	&	(0.005\%)	&	(15.6\%)	&	(0.19\%)	&	(4.5\%)	\\
1.1	&	2.35	&	0.8251	&	0.8252	&	0.8252	&	0.8251	&	0.8251	&	0.8251	&	0.7673	&	0.8255	&	0.8170	\\
	&		&	(0.18\%)	&	(0.006\%)	&	(0.003\%)	&	(0.005\%)	&	(0.005\%)	&	(0.005\%)	&	(7.3\%)	&	(0.04\%)	&	(1.0\%)	\\
\hline
\end{tabular}
\centering
\label{tbl:Ni56 Sens Z0}
\end{minipage}
\end{table*}

\begin{table*}
\begin{minipage}{160mm}
\caption{Same as Table~\ref{tbl:t0 Sens Z1} for $Z=0$.}
\begin{tabular}{|c||c||c||c||c||c||c||c||c||c||c|}
\hline
$\MWD\,[M_{\odot}]$  & $\Delta x_0\,[\textrm{km}]$ & Reference & NSE6  & NSE7 & w/o ASE & w weak & w thermal $\nu$ & w/o Coul. & w/o ex. & w/o screen \\ \hline
0.8	&	1.75	&	32.69	&	32.68	&	32.69	&	32.68	&	32.68	&	32.69	&	33.66	&	32.69	&	32.56	\\
	&		&	(0.97\%)	&	(0.003\%)	&	(0.000\%)	&	(0.001\%)	&	(0.007\%)	&	(0.001\%)	&	(2.9\%)	&	(0.01\%)	&	(0.4\%)	\\
0.85	&	0.83	&	33.24	&	33.24	&	33.24	&	33.24	&	33.24	&	33.24	&	33.66	&	33.21	&	33.33	\\
	&		&	(0.06\%)	&	(0.000\%)	&	(0.000\%)	&	(0.002\%)	&	(0.010\%)	&	(0.002\%)	&	(1.2\%)	&	(0.07\%)	&	(0.3\%)	\\
0.9	&	1.56	&	32.46	&	32.46	&	32.46	&	32.46	&	32.45	&	32.46	&	33.53	&	32.44	&	32.63	\\
	&		&	(0.25\%)	&	(0.000\%)	&	(0.001\%)	&	(0.002\%)	&	(0.017\%)	&	(0.000\%)	&	(3.2\%)	&	(0.05\%)	&	(0.5\%)	\\
1	&	2.74	&	31.11	&	31.11	&	31.11	&	31.11	&	31.10	&	31.11	&	31.97	&	31.11	&	31.26	\\
	&		&	(0.08\%)	&	(0.005\%)	&	(0.006\%)	&	(0.008\%)	&	(0.034\%)	&	(0.003\%)	&	(2.7\%)	&	(0.01\%)	&	(0.5\%)	\\
1.1	&	2.35	&	30.47	&	30.47	&	30.47	&	30.47	&	30.46	&	30.47	&	31.09	&	30.48	&	30.49	\\
	&		&	(0.15\%)	&	(0.005\%)	&	(0.005\%)	&	(0.004\%)	&	(0.042\%)	&	(0.008\%)	&	(2.0\%)	&	(0.03\%)	&	(0.1\%)	\\
\hline
\end{tabular}
\centering
\label{tbl:t0 Sens Z0}
\end{minipage}
\end{table*}

\begin{table*}
\begin{minipage}{160mm}
\caption{Same as Table~\ref{tbl:Ni56 Sens Z1 FLASH} for $Z=0$.}
\begin{tabular}{|c||c||c||c||c||c||c|}
\hline
$\MWD\,[M_{\odot}]$  & $\Delta x\,[\textrm{km}]$ & $lr_{\max}$  & $\nimass\,[M_{\odot}]$  &  $\nimass\,[M_{\odot}]$   &  $t_0\,[\textrm{day}]$ &  $t_0\,[\textrm{day}]$    \\
 & &  & Reference & w/o ASE & Reference & w/o ASE  \\ \hline
0.8	&	0.25	&	16	&	0.055	&	0.055	&	32.8	&	32.8	\\
     	&		&		&	(3.90\%)	&	(0.071\%)	&	(0.05\%)	&	(0.013\%)	\\
     0.85	&	0.25	&	16	&	0.147	&	0.146	&	33.2	&	33.1	\\
     	&		&		&	(2.13\%)	&	(1.021\%)	&	(0.03\%)	&	(0.057\%)	\\
     0.9	&	0.5	&	15	&	0.281	&	0.277	&	32.4	&	32.4	\\
     	&		&		&	(2.26\%)	&	(1.437\%)	&	(0.18\%)	&	(0.047\%)	\\
     1	&	2	&	13	&	0.562	&	0.556	&	31.2	&	31.2	\\
     	&		&		&	(2.19\%)	&	(0.931\%)	&	(0.42\%)	&	(0.155\%)	\\
     1.1	&	4	&	12	&	0.808	&	0.804	&	30.7	&	30.8	\\
     	&		&		&	(2.62\%)	&	(0.487\%)	&	(1.11\%)	&	(0.134\%)	\\
\hline
\end{tabular}
\centering
\label{tbl:Ni56 Sens Z0 FLASH}
\end{minipage}
\end{table*}

\section{Full radiation transfer calculations}
\label{sec:full RT}

Full radiation transfer is calculated using the Monte Carlo code URILIGHT \citep{Wygoda2019a}. For a detailed description of the code, and its verification through comparison to other codes, see \citep{Wygoda2019b}. For the radiation transfer calculations, the hydrodynamical grid from V1D or FLASH is remapped onto a $100$-cell, uniform mass grid, and a logarithmically spaced time grid of $128$ steps between $2$ day and $210$ day (the simulations were checked for convergence with respect to spatial and time resolutions). Atomic line data for the bound-bound transitions, which constitute the main source of opacity for the computation of the light curves, are taken from \cite{Kurucz1} and \cite{Kurucz23}. The probability of absorption was set to be $\epsilon=0.8$ \citep{Kasen2006}. The radioactive decay chains of $^{37}$K, $^{48}$Cr, $^{49}$Cr, $^{51}$Mn, $^{52}$Fe, $^{55}$Co, $^{56}$Ni and $^{57}$Ni are included. 

\begin{figure}
	\includegraphics[width=0.48\textwidth]{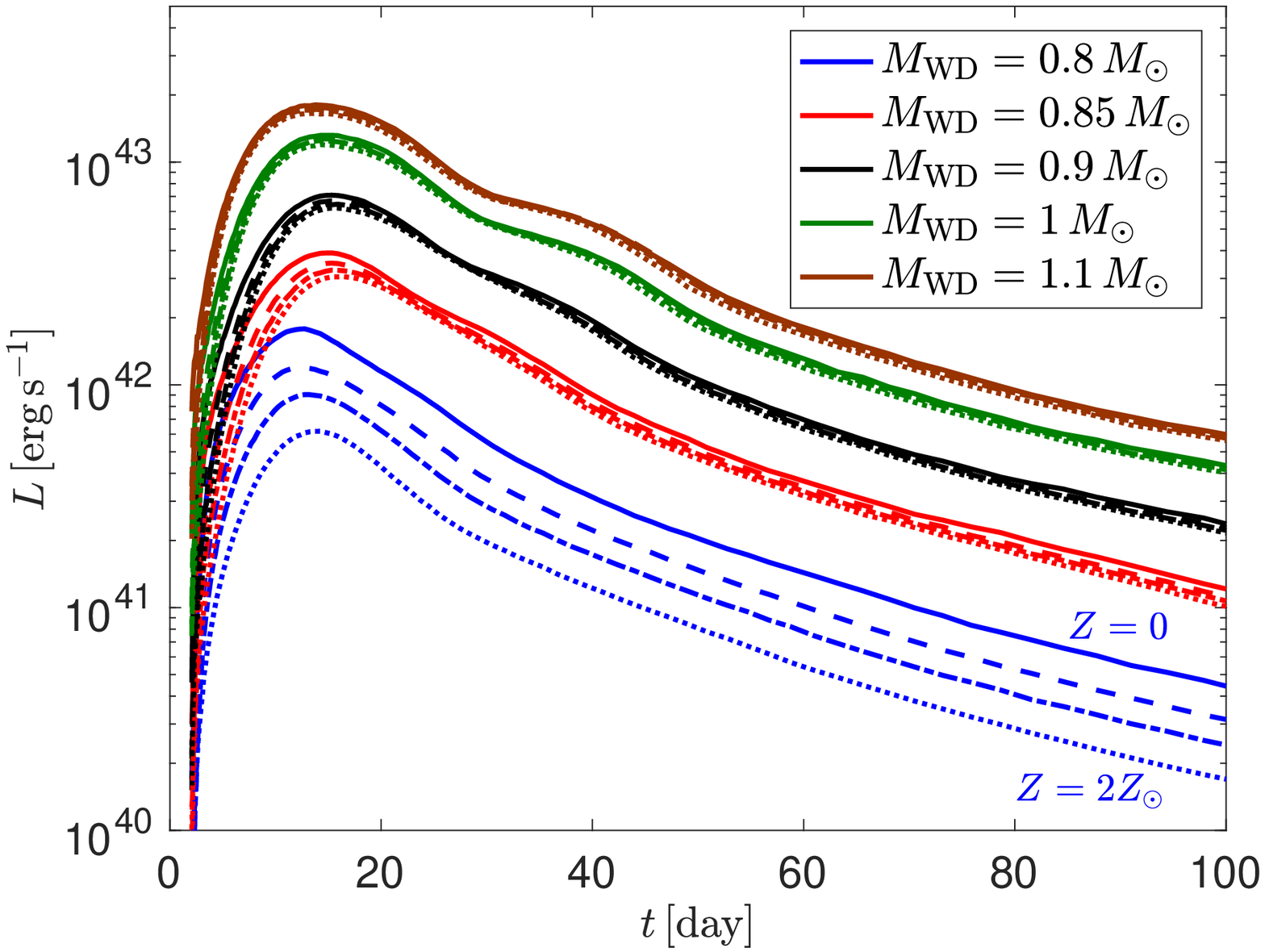}
	\caption{Bolometric light curves as a function of time since explosion. The light curves are calculated with the Monte Carlo code URILIGHT for the V1D converged ejecta of $\MWD=0.8,0.85,0.9,1,1.1 M_{\odot}$ (blue, red, black, green and brown lines, respectively) and $Z=0,0.5,1,2 Z_{\odot}$ (solid, dashed, dash-dotted, and dotted lines, respectively).
		\label{fig:bolos}}
\end{figure}

The calculated bolometric light curves for the converged V1D ejecta are presented in Figure~\ref{fig:bolos} (the results for the converged FLASH ejecta are very similar). As can be seen in the figure, the metallicity mostly affects the results of the low luminosity light curves (in agreement with the results of Section~\ref{sec:results}). We use the same methods as those of \citet{Sharon2020} to extract the $\gamma$-ray deposition history from the bolometric light curve. For this procedure, we need to define a time range, $t_{L=Q}$, over which the bolometric luminosity, $L$, equals the energy deposition rate, $Q$ \citep[see detailed discussion in][]{Sharon2020}. While it is straight forward to define such a time range for observed SNe Ia, there is an unrealistic recombination to neutral iron obtained at $\mysim100\,\textrm{day}$ in the radiation transfer simulations, which leads to deviations from the assumption $L=Q$. In order to bypass this problem, we use the calculated $Q$ and define $t_{L=Q}$ as times when the deviation between $L$ and $Q$ is smaller than a few percent. From this process, we obtain $t_{0}^{\rm{RT}}$ values, which are presented in Tables~\ref{tbl:Ni56 Conv Z1}, \ref{tbl:Ni56 Conv Z0}, \ref{tbl:Ni56 Conv Z05}, and~\ref{tbl:Ni56 Conv Z2}. We also calculate the $\gamma$-ray deposition history by including the recombination to neutral iron within $t_{L=Q}$. The deviation from the previous analysis is smaller than $1.5\,\textrm{day}$ in all cases, so we assign an uncertainty of $\mysim1\,\textrm{day}$ to $t_{0}^{\rm{RT}}$. 

\end{appendix}

\bsp	
\label{lastpage}
\end{document}